\documentclass[]{IEEEtran}
\usepackage{amsfonts}
\IEEEoverridecommandlockouts

\ifCLASSINFOpdf
\else
\fi
\usepackage{epstopdf}
\usepackage{multirow}
\usepackage{cite}
\usepackage{stfloats}
\usepackage{color}
\usepackage{epsfig}
\usepackage{graphicx}
\usepackage{psfig}
\usepackage{epsf}
\usepackage{slashbox}
\usepackage{float}
\usepackage{amssymb }
\usepackage[cmex10]{amsmath}
\usepackage{algorithm} 
\usepackage{algorithmic}
\usepackage{setspace}
\usepackage{amsfonts}
\usepackage{amsmath}
\usepackage{booktabs}
\usepackage{amsmath,bm}
\usepackage[justification=centering]{caption}
\begin{document}
\title{Relay-Enabled Backscatter Communications: Linear Mapping and Resource Allocation}
\author{Rui Xu, Liqin Shi, Yinghui Ye, Haijian Sun, and Gan Zheng, \emph{Fellow, IEEE}\vspace{-15pt}

\thanks{Rui Xu, Liqin Shi, and Yinghui Ye are with the Shaanxi Key Laboratory of
Information Communication Network and Security, Xi'an University of Posts
\& Telecommunications, China (e-mails: dhlscxr@126.com, liqinshi@hotmail.com, connectyyh@126.com).}
\thanks{Haijian Sun is with  the School of Electrical and Computer
Engineering, The University of Georgia, Athens, GA, USA (e-mail: hsun@uga.edu). }
\thanks{Gan Zheng is with the School of Engineering, University of Warwick,
Coventry, CV4 7AL, UK (email: Gan.Zheng@warwick.ac.uk).}
}
\markboth{}
{Shi\MakeLowercase{\textit{et al.}}:}
\maketitle

\begin{abstract}
Relay-enabled backscatter communication (BC) is an intriguing paradigm to alleviate
energy shortage and improve throughput of Internet-of-Things (IoT) devices.
Most of the existing works focus on the resource allocation that considered the unequal and continuous time allocation for both source-relay and relay-destination links.
However, the continuous time allocation
 may be infeasible since in practice, the time allocation shall be carried out in integral multiple of the subframe duration unit.
In this article, we study a discrete time scheme from the perspective of frame structure, where one transmission block is divided into two phases and the linear mapping is employed as a re-encoding method to determine the number of subframes for both phases and the power allocation for each subframe in a relay-enabled BC system.
Based on this, we derive an accurate system-throughput expression and formulate a mixed-integral non-convex optimization problem to maximize the system throughput by jointly optimizing the power reflection coefficient
(PRC) of the IoT node, the power allocation of the hybrid access point (HAP) and the linear mapping matrix, and solve it via a three-step approach. First, we derive the optimal eigenvalues of the linear mapping matrix via introducing auxiliary variables and exploiting Karush-Kuhn-Tucker (KKT) conditions. Second, we employ the time-sharing technology to convert the discrete optimization problem to a continuous non-convex problem and solve it by leveraging successive convex approximation (SCA) and monotonicity or bisection method.
Finally, an integer conversion strategy is designed to find a suboptimal number of subframes for both phases, based on which the continuous variables are further optimized. According to the above approaches, we propose a low complexity iterative algorithm to obtain the throughput maximization-based resource allocation solution. Numerical results analyze the performance of our proposed algorithm, verify the superiority of our proposed scheme, and evaluate the impacts of network parameters on the system throughput.
\end{abstract}
\IEEEpeerreviewmaketitle
\section{Introduction}
As an integral part of the future Internet, the Internet of Things (IoT) is expected to promote the transformation of human production and lifestyle in a fully intelligent and automated way \cite{9509294,7123563}.
Towards this goal, it is required to realize high-density IoT
deployment and  seamless connectivity among IoT nodes. However, due to the limitations of battery size and device cost, most IoT nodes are energy-constrained and may run out of battery in a few months or years, leading to an outage event and a threat to  seamless connectivity \cite{9837295,9051982}.
To alleviate the challenge, wireless power transfer has been incorporated into wireless communications, leading to wireless powered communication network (WPCN) \cite{7984754} and simultaneous wireless information and power transfer (SWIPT) \cite{8214104}. Both solutions allow IoT nodes to harvest energy from the radio frequency signals transmitted by the dedicated energy source, e.g., power beacon (PB), and utilize it to support active communications (AC).  Nevertheless, due to the contradiction between the low energy harvesting (EH) efficiency and power-hungry AC, the harvested energy may not be enough to support AC adopted in WPCN and SWIPT, which therefore motivates us to find a ultra-low power communication paradigm, i.e., backscatter communication (BC) \cite{8368232,8999426}. Different from AC, in BC, IoT nodes (also referred to as BC transmitter in the following context) are allowed to passively modulate and reflect the incident signal by adjusting the antenna load impedance so as to avoid using active components, and simultaneously to harvest energy for sustaining its circuit operation. Accordingly, in BC, the IoT node has a simpler tranceiver diagram and lower power consumption.

Despite these advantages of BC, it is appropriate for short-range communications \cite{9837295} and cannot meet the requirement of long-range IoT applications. One promising way to extend the communication range is the hybrid paradigm of BC and AC, where the IoT nodes can either modulate and backscatter their own information to the receiver via BC, or transmit information after harvesting enough energy, subject to the energy-causality constraint of each IoT node. The hybrid architecture can fully exploit  the complementary capability of BC and AC in terms of the power consumption and throughput to obtain a better performance, as verified by\cite{9159908,9161012,9449060}. However, due to the low efficiency of EH, the IoT node has to allocate a long period for EH and only leave a very short interval for AC in each transmission block, thus hindering its practicability. Another alternative solution is the relay-enabled BC$\footnote{We note that there exists another paradigm combining BC with relaying (see  \cite{8302460,8885856,8647737,8746225,8746689}), which is different from our work and may not be used to extend BC's  communication range. The reasons can be summarized as follows. In  \cite{8302460,8885856,8647737,8746225,8746689}, the relay is energy-constrained and operates in the BC by  reflecting the source's signal without modulating its own information, while in our work, the relay works in the AC mode to forward the BC transmitter's signal.}$,
where a relay is deployed to forward BC transmitter's signal \cite{9860432,8417655,8490651,8957679,8943100,9082191,9473528}. In particular, the IoT node modulates and backscatters incident signals from a PB to the relay and the associated receiver simultaneously, and subsequently the relay decodes the received signal and forwards the decoded signal to the receiver. By doing so, both source-destination and source-relay-destination links form a virtual multi-antenna scenario that exploits the radio diversity gain to improve the performance and thus to extend the communication range.

In \cite{9860432}, the authors formulated a throughput maximization problem by jointly optimizing the power reflection coefficient (PRC) of the IoT node and the operation modes of the hybrid access point (HAP), i.e., the relay or the PB, based on the equal-time allocation in an opportunistic relay-enabled BC system.
However, the equal-time allocation may not make full use of the transmission for a better performance.
Therefore, the resource allocation based on the unequal-time allocation has become an important research direction in relay-enabled BC.
In \cite{8417655}, the authors maximized system throughput of the relay-enabled BC by optimizing the BC time of the IoT node, the AC time and EH time of the relay when the relay is energy-constrained, or optimizing time allocation without EH when the relay is non-energy-constrained.
Considering the same scenario with \cite{8417655}, the authors in \cite{8490651} studied the transmission time minimization problem with a given amount of information required to be delivered.
Subsequently, in \cite{8957679} and \cite{8943100}, the authors extended the single relay scenario \cite{8417655,8490651} into the multiple relays scenario in the absence of direct link between the IoT node and the destination.
Under this setting, a sum-rate maximization problem was solved in \cite{8957679} to obtain the optimal time allocated for the IoT node to perform BC, and for the multiple relays to perform AC and EH, while the authors in \cite{8943100} formulated a sum-rate maximization problem by jointly optimizing the energy beamforming vectors of the HAP, the time scheduling among EH, BC, AC, as well as the power allocation of the relays.
Nevertheless, the above works \cite{8417655,8490651,8957679,8943100} have not considered the energy-causality constraints of the IoT nodes and may not work well in the scenario with energy-constrained IoT nodes.
To fill this gap, the authors in \cite{9082191} studied a throughput maximization problem by optimizing time and power allocation subject to the energy-causality constraints of the IoT node and the relay.
Considering the scenario with direct link and the energy-causality constraint of the IoT node, \cite{9473528} proposed an optimal time allocation strategy to maximize the achievable throughput.

Although \cite{8417655,8490651,8957679,8943100,9082191,9473528}  have designed various resource allocation schemes for relay-enabled BC systems by considering unequal and continuous time allocation, there are two drawbacks.
First, the unequal and continuous time allocation may be impractical in relay-enabled BC systems.
In practice, the transmission block is composed of frames and each frame is divided into a number of subframes with fixed duration rather than arbitrary segmentation, thus the practical unequal time allocation shall be carried out in integral multiple of the subframe duration unit. This indicates that the time allocated for both phases in relay-enabled BC systems should be discrete instead of continuous one.
Second, when the direct link exists, it is quite hard to derive an accurate throughput for the
relay-enabled BC system. To address this issue, the existing works \cite{8417655,8490651,9473528} adopted an upper bound of the system throughput\cite{7274646}, but such a way leads to the overestimation of the performance.
 Based on the above discussion, it is necessary to revisit the throughput expression and redesign the resource
allocation scheme in relay-enabled BC systems while considering the unequal and discrete time allocation.

Towards this end, in this paper, we consider a relay-enabled BC network in the presence of a direct link, where a HAP serves as the PB to broadcast an energy signal and an IoT node conveys information to the HAP and destination via BC in the first phase, and then the HAP serves as the relay to re-encode the information of the IoT node and forward it to the destination in the second phase.
To achieve the unequal and discrete time allocation for both phases, the linear mapping\footnote{The linear mapping is implemented through a linear mapping matrix, in  which the dimensions and eigenvalues of the linear mapping matrix reflect the number of subframes allocated for the two phases and the power allocated for each subframe, respectively.} \cite{6784109,6884029,7795362,4663931}
is introduced as the re-encoding method\footnote{The re-encoding method can be implemented by using the linear mapping to transform the decoded original signal into the forwarded signal at the relay, which ensures that the dimensions of the forwarded signal can adapt to the number of subframes allocated to
the relay phase.} through a linear mapping matrix,
  which can adjust the number of subframes for both phases and allocate the power for each subframe dynamically.
Under this setting, we aim to design a practical resource allocation scheme to maximize the throughput of the considered relay-enabled BC network. Our main contributions are summarized below:
\begin{itemize}
  \item This is the first work to consider the unequal and discrete time allocation from the perspective of frame structure in a relay-enabled BC network. We derive an accurate system-throughput expression for the relay-enabled BC network by introducing the linear mapping as the re-encoding method and combining the received signal at the receiver to jointly detect the original signal from the perspective of the frame, avoiding the overestimation of the performance caused by the inaccurate throughput in \cite{8417655,8490651,8957679,8943100,9082191,9473528}.
  \item We study the throughput maximization for a relay-enabled BC network under the unequal and discrete time allocation scheme.
 More specifically, we propose a mixed-integer non-convex optimization problem to maximize the system throughput by jointly optimizing the linear mapping matrix, the power allocation of the HAP and the PRC of the IoT node, subject to the energy-causality constraint of the IoT node, the energy constraint of the HAP and the constraint of the linear mapping matrix. The formulated problem is different from the existing works in that we not only introduce a linear mapping matrix of the relay, but also apply our derived accurate throughput, while these have not been considered \cite{8417655,8490651,8957679,8943100,9082191,9473528}.
 Due to this, our formulated problem is much more challenging than the existing ones.
  \item To solve the mixed-integer non-convex optimization problem, we derive the optimal eigenvalues of the linear mapping matrix via introducing auxiliary variables and exploiting Karush-Kuhn-Tucker (KKT) conditions. Then, we apply the time-sharing technology to convert the discrete optimization problem to a continuous non-convex one that can be cast into two problems: one is solved by leveraging successive convex approximation (SCA) and the other is solved by exploiting the monotonicity or bisection method. Since the continuous time allocation is obtained, we design an integer conversion strategy to serialize the continuous optimal results into discrete values that satisfy the original constraints, based on
which the continuous variables are further optimized. On this basis, we propose an iterative algorithm to maximize the throughput of our considered network.
  \item We perform extensive simulations to verify the convergence of our proposed iterative algorithm, analyze the gap caused by the time-sharing technology, and evaluate the throughput achieved by our proposed algorithm and the impact of network parameters on throughput, resulting in the following conclusions. 1) The proposed SCA-based iterative algorithm has a fast convergence speed; 2) The gap caused by the time-sharing technology decreases with the increase of the total number of subframes in one transmission block; 3) The proposed scheme shows superiority comparing with other schemes; 4) To maximize the throughput, more subframes are allocated to the relay phase when the channel condition of the source-destination link is worse.
\end{itemize}

The rest of this paper is organized as follows. In section II, we describe the system model and analyze the working flow of the relay-enabled BC network. Section III studies the throughput maximization problem and gives the transformations and solutions.
Section IV provides simulation results to validate the effectiveness of the proposed algorithms and the superiority of the proposed scheme. Finally, Section V draws the conclusion of our work.

\section{System Model}
As depicted in Fig. 1, we consider a relay-enabled BC system consisting of one IoT node (also termed as S), one receiver (also termed as D) and one HAP (also termed as R) with two functions of the relay and PB. Both the IoT node and receiver are equipped with a single antenna, while the HAP has two antennas so that it can broadcast energy signals and receive the backscattered signal from the IoT node simultaneously. In the considered system, we assume that one transmission block is composed of $L$ subframes and its duration time is denoted by $T_s$. There are two transmission phases in one transmission block, i.e., the backscatter phase with $M$ subframes and the relay phase with $N$ subframes, and the number of subframes for two phases satisfies $L=M+N$. In the backscatter phase, the HAP broadcasts energy signals and the IoT node modulates and reflects its own information
 on a part of incident signals, and harvests energy from the remaining incident signals to support its circuit operation, simultaneously.
In the relay phase, the HAP, serving as a relay, re-encodes the decoded signal and forwards it to the receiver, while the IoT node keeps silent. Such an assumption of the IoT node avoids the simultaneous reception at the receiver, and makes the network easy to implement.
We suppose that the duration time of one transmission block, $T_s$, is less than the coherence time, i.e., the channel power gain of all links keep unchanged within each transmission block. We also denote the channel coefficients, i.e., S$-$D link, S$-$R link and R$-$D link, as ${h_{{\rm{SD}}}}$, ${h_{{\rm{SR}}}}$, and ${h_{{\rm{RD}}}}$, respectively. For traceability analysis, we stack all subframes together and re-write the entire transmission block in a matrix format, thus the channel matrices of S$-$D, S$-$R and R$-$D links can be denoted as ${{\bf{H}}_{{\rm{SD}}}} = {h_{{\rm{SD}}}}{{\bf{I}}_M}$, ${{\bf{H}}_{{\rm{SR}}}} = {h_{{\rm{SR}}}}{{\bf{I}}_M}$, and ${{\bf{H}}_{{\rm{RD}}}} = {h_{{\rm{RD}}}}{{\bf{I}}_N}$, where ${{\bf{I}}_M}$ and ${{\bf{I}}_N}$ are the identity matrices with $M$ and $N$ dimensions, respectively.
\begin{figure}
  \centering
  \includegraphics[width=0.35\textwidth]{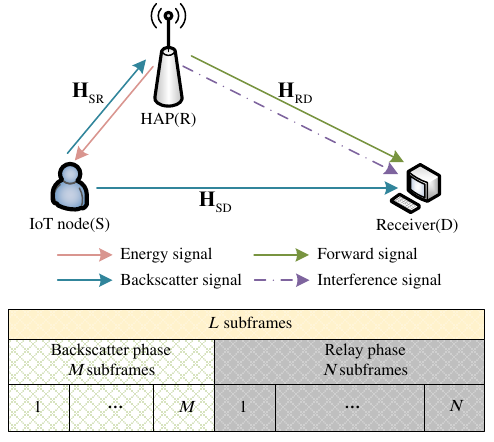}  \\
  \caption*{Fig. 1. \small{System model and frame structure.}}\label{fig1}
\end{figure}
\subsubsection{The backscatter phase}
During the backscatter phase, the HAP broadcasts the energy signal, then the received signal of the IoT node is given by
\begin{align}\label{1}
{\bf{y}} = \sqrt {{P_0}} {{\bf{H}}_{{\rm{SR}}}}{\bf{s}},
\end{align}
where ${\bf{s}} \in {\mathbb{C}^{M \times 1}}$ is the predefined energy signal with unit power and $P_0$ denotes the transmit power of the HAP. Note that in \eqref{1}, the noise of the IoT node is neglected since its circuit only consists passive components and takes few signal processing operations \cite{7551180,8424210,7769255}.

By performing BC, the received signal of the IoT node ${\bf{y}}$ is split into two parts via a PRC $\beta \left( {0 \le \beta  \le 1} \right)$. The IoT node modulates its information on one part of the received signal
 and reflects to the receiver and the HAP, and the remaining part is absorbed by the energy harvester. Thus, the backscattered signal and the harvested energy of the IoT node can be written as \cite{4447346,4447347,7556997,6685977,9358202}, respectively,
\begin{align}\label{2}
{{\bf{y}}_{\rm{B}}} = \sqrt {{P_{\rm{0}}}\beta } {\bf{F}}{\bf{c}},
\end{align}
\begin{align}\label{3}
{E_{\rm{h}}}{\rm{ = }}\frac{{{T_s}\eta \!\left( {1 \!\! -\!\!\beta} \right)\!{P_0}}}{L}{\rm{Tr}}\left(\! {{{\bf{H}}_{{\rm{SR}}}}\!{\bf{H}}_{{\rm{SR}}}^{\cal H}} \!\right){\rm{ \!= }}\frac{M}{L}\!{T_s}\eta \! \left( {1\!\! - \!\!\beta } \right)\!{P_0}{\left| {{h_{{\rm{SR}}}}} \right|^2},
\end{align}
where ${\bf{c}} \in {\mathbb{C}^{M \times 1}}$ denotes the information of the IoT node with unit power, ${\bf{F}} = {\rm{diag}}\left( {{{\bf{H}}_{{\rm{SR}}}}{\bf{s}}} \right)$ and $\eta$ is the energy conversion efficiency$\footnote{For ease of analysis, we consider a linear energy harvesting model. Please note that our work can be easily extended to the  scenario with non-linear model scenario by following a similar approach considered in \cite{9866050} and \cite{8675436}.}$ of the EH circuit.

Then, the received backscatter signal at the receiver and HAP can be expressed as, respectively,
\begin{align}\label{4}
{{\bf{y}}_{{\rm{SD}}}} = \sqrt {{P_0}\beta } {{\bf{H}}_{{\rm{SD}}}}{\bf{Fc}} + \sqrt {{P_{\rm{0}}}} {{\bf{H}}_{{\rm{RD}}}}{\bf{s}}{\rm{ + }}{{\bf{n}}_{{\rm{SD}}}},
\end{align}
\begin{align}\label{5}
{{\bf{y}}_{{\rm{SR}}}} = \sqrt {{P_{\rm{0}}}\beta } {{\bf{H}}_{{\rm{SR}}}}{\bf{Fc}} + \sqrt {{P_{\rm{0}}}} {{\bf{H}}_{{\rm{LI}}}}{\bf{s}} + {{\bf{n}}_{{\rm{SR}}}},
\end{align}
where ${{\bf{n}}_{{\rm{SD}}}} \in {\mathbb{C}^{M \times 1}}$ and ${{\bf{n}}_{{\rm{SR}}}} \in {\mathbb{C}^{M \times 1}}$ are the Gaussian noise vectors,  ${{\bf{H}}_{\rm{LI}}} \in {\mathbb{C}^{M \times M}}$ denotes the residual loop interference channel gain matrix at the HAP, and ${{\bf{H}}_{{\rm{LI}}}} = {h_{{\rm{LI}}}}{{\bf{I}}_M}$ with the unit matrix of $M$ dimensions ${{\bf{I}}_M}$.

Eqs. \eqref{4} and \eqref{5} indicate that there exists co-channel interference caused by the energy signal. Since the energy signal is predefined and the channel state information (CSI) is assumed to be available, the co-channel interference, i.e., $\sqrt {{P_{\rm{0}}}} {{\bf{H}}_{{\rm{RD}}}}{\bf{s}}$ and $\sqrt {{P_{\rm{0}}}} {{\bf{H}}_{{\rm{LI}}}}{\bf{s}}$, can be removed via successive interference cancellation (SIC) before decoding $\bf{c}$ at both the HAP and receiver.
Then using Shannon capacity, we can write the throughput of the S$-$D and S$-$R links as, respectively,
\begin{align}\notag
{R_{{\rm{SD}}}} &= \frac{{{T_s}W}}{L}{\log _2}\left[ {\det \left( {{{\bf{I}}_M} + \frac{{\beta {P_0}{{\bf{H}}_{{\rm{SD}}}}{\bf{Q}}{{\bf{Q}}^{\cal H}}{\bf{H}}_{{\rm{SD}}}^{\cal H}}}{{W{\sigma ^2}}}} \right)} \right]\\
 &= \frac{M}{L}{T_s}W{\log _2}\left( {1 + \frac{{\beta {P_0}{{{\left| {{h_{{\rm{SR}}}}} \right|}^2}{{\left| {{h_{{\rm{SD}}}}} \right|}^2}} }}{{W{\sigma ^2}}}} \right),\label{8}
\end{align}
\begin{align}\notag
{R_{{\rm{SR}}}}& = \frac{{{T_s}W}}{L}{\log _2}\left[ {\det \left( {{{\bf{I}}_M} + \frac{{\beta {P_0}{{\bf{H}}_{{\rm{SR}}}}{\bf{Q}}{{\bf{Q}}^{\cal H}}{\bf{H}}_{{\rm{SR}}}^{\cal H}}}{{W{\sigma ^2}}}} \right)} \right]\\
 &= \frac{M}{L}{T_s}W{\log _2}\left( {1 + \frac{{\beta {P_0}{{\left| {{h_{{\rm{SR}}}}} \right|}^4}}}{{W{\sigma ^2}}}} \right),\label{9}
\end{align}
where $W$ is the communication bandwidth and ${\sigma ^2}$ denotes the noise power spectrum density. In \eqref{8} and \eqref{9}, ${\bf{Q}} = {\rm{diag}}\left( {{{\bf{H}}_{{\rm{SR}}}}\bf{1}} \right)$, where $\bf{1}$ is an $M \times 1$ column vector
whose entries are all one \cite{4447346,4447347,7556997,6685977,9358202}.

\subsubsection{The relay phase}
In this phase, the IoT node keeps silent and the HAP forwards the received signal to the receiver following decode-and-forward (DF) protocol. Thus, the received signal of the receiver is given by
\begin{align}\label{10}
{{\bf{y}}_{{\rm{RD}}}} = \sqrt {{P_1}} {{\bf{H}}_{{\rm{RD}}}}{\bf{x}}{\rm{ + }}{{\bf{n}}_{{\rm{RD}}}}= \sqrt {{P_1}} {{\bf{H}}_{{\rm{RD}}}}{\bf{Gc}}{\rm{ + }}{{\bf{n}}_{{\rm{RD}}}},
\end{align}
where $P_1$ represents the forwarding power of the HAP, ${\bf{x}} \in {\mathbb{C}^{N \times 1}}$  is the re-encoded signal with unit average power via linear mapping method, ${\bf{G}} \in {\mathbb{C}^{N \times M}}$ is the mapping matrix to transform the original signal into the re-encoded signal, i.e., ${\bf{x}} = {\bf{Gc}}$, and ${{\bf{n}}_{{\rm{RD}}}} \in {\mathbb{C}^{N \times 1}}$ is the Gaussian noise vector.

At the receiver, it combines the received signal ${{\bf{y}}_{{\rm{SD}}}}$ and ${{\bf{y}}_{{\rm{RD}}}}$ to jointly detect the information $\bf{c}$.
Then the combined signal is given by
\begin{align}\label{11}
{{\bf{y}}_{\rm{D}}} \buildrel \Delta \over = \left( {\begin{array}{*{20}{c}}
{{{\bf{y}}_{{\rm{SD}}}}}\\
{{{\bf{y}}_{{\rm{RD}}}}}
\end{array}} \right) = \left( {\begin{array}{*{20}{c}}
{\sqrt {{P_0}\beta } {{\bf{H}}_{{\rm{SD}}}}{\bf{Fc}}}\\
{\sqrt {{P_1}} {{\bf{H}}_{{\rm{RD}}}}{\bf{Gc}}}
\end{array}} \right) + \left( {\begin{array}{*{20}{c}}
{{{\bf{n}}_{{\rm{SD}}}}}\\
{{{\bf{n}}_{{\rm{RD}}}}}
\end{array}} \right).
\end{align}
It can be further simplified as
\begin{align}\label{12}
{{\bf{y}}_{\rm{D}}} = {{\bf{H}}_{\rm{D}}}{\bf{c}} + {{\bf{n}}_{\rm{D}}},
\end{align}
where ${{\bf{H}}_{\rm{D}}} \buildrel \Delta \over = \left( {\begin{array}{*{20}{c}}
{\sqrt {{P_0}\beta } {{\bf{H}}_{{\rm{SD}}}}{\bf{F}}}\\
{\sqrt {{P_1}} {{\bf{H}}_{{\rm{RD}}}}{\bf{G}}}
\end{array}} \right)$ and ${{\bf{n}}_{\rm{D}}} \buildrel \Delta \over = \left( {\begin{array}{*{20}{c}}
{{{\bf{n}}_{{\rm{SD}}}}}\\
{{{\bf{n}}_{{\rm{RD}}}}}
\end{array}} \right)$.

Using \eqref{12}, the expression of achievable throughput at the receiver and the transformation can be written as \eqref{13}, as shown at the top of the next page.
\begin{figure*}
\begin{align}\notag
{R_{\rm{D}}} &= \frac{{{T_s}W}}{L}{\log _2}\left[ {\det \left( {{{\bf{I}}_L} + \frac{{{{\bf{H}}_{\rm{D}}}{\bf{H}}_{\rm{D}}^{\cal H}}}{{W{\sigma ^2}}}} \right)} \right] = \frac{{{T_s}W}}{L}{\log _2}\left[ {\det \left( {{{\bf{I}}_M} + \frac{{{\bf{H}}_{\rm{D}}^{\cal H}{{\bf{H}}_{\rm{D}}}}}{{W{\sigma ^2}}}} \right)} \right]\\\notag
& = \frac{{{T_s}W}}{L}{\log _2}\left[ {\det \left( {{{\bf{I}}_M} + \frac{{\beta {P_0}{{\bf{H}}_{{\rm{SD}}}}{\bf{Q}}{{\bf{Q}}^{\cal H}}{\bf{H}}_{{\rm{SD}}}^{\cal H}}}{{W{\sigma ^2}}} + \frac{{{P_1}{\bf{H}}_{{\rm{RD}}}^{\cal H}{{\bf{G}}^{\cal H}}{\bf{G}}{{\bf{H}}_{{\rm{RD}}}}}}{{W{\sigma ^2}}}} \right)} \right]\\\notag
& = \frac{{{T_s}W}}{L}{\log _2}\left[ {\det \left( {{{\bf{I}}_M} + \frac{{\beta {P_0}{{\left| {{h_{{\rm{SR}}}}} \right|}^2}{{\left| {{h_{{\rm{SD}}}}} \right|}^2}}}{{W{\sigma ^2}}}{{\bf{I}}_M} + \frac{{{P_1}{{\left| {{h_{{\rm{RD}}}}} \right|}^2}}}{{W{\sigma ^2}}}{{\bf{G}}^{\cal H}}{\bf{G}}} \right)} \right]\\\notag
 &= \frac{{{T_s}W}}{L}{\log _2}\left[ {\det \left( {\left( {1 + \frac{{\beta {P_0}{{\left| {{h_{{\rm{SR}}}}} \right|}^2}{{\left| {{h_{{\rm{SD}}}}} \right|}^2}}}{{W{\sigma ^2}}}} \right){{\bf{I}}_M} + \frac{{{P_1}{{\left| {{h_{{\rm{RD}}}}} \right|}^2}}}{{W{\sigma ^2}}}{{\bf{G}}^{\cal H}}{\bf{G}}} \right)} \right]\\\notag
 &= \frac{{{T_s}W}}{L}{\log _2}\left[ {{{\left( {1 + \frac{{\beta {P_0}{{\left| {{h_{{\rm{SR}}}}} \right|}^2}{{\left| {{h_{{\rm{SD}}}}} \right|}^2}}}{{W{\sigma ^2}}}} \right)}^M} \times \det \left( {{{\bf{I}}_N} + \frac{{{P_1}{{\left| {{h_{{\rm{RD}}}}} \right|}^2}/\left({W\sigma ^2}\right)}}{{1 + \beta {P_0}{{\left| {{h_{{\rm{SR}}}}} \right|}^2}{{\left| {{h_{{\rm{SD}}}}} \right|}^2}/\left({W\sigma ^2}\right)}}{\bf{G}}{{\bf{G}}^{\cal H}}} \right)} \right]\\\label{13}
 &= \underbrace {\frac{{{T_s}WM}}{L}{{\log }_2}\left( {1 + \frac{{\beta {P_0}{{\left| {{h_{{\rm{SR}}}}} \right|}^2}{{\left| {{h_{{\rm{SD}}}}} \right|}^2}}}{{W{\sigma ^2}}}} \right)}_{{R_{{\rm{SD}}}}} + \underbrace {\frac{{{T_s}W}}{L}{{\log }_2}\left[ {\det \left( {{{\bf{I}}_N} + \frac{{{P_1}{{\left| {{h_{{\rm{RD}}}}} \right|}^2}/\left({W\sigma ^2}\right)}}{{1 + \beta {P_0}{{\left| {{h_{{\rm{SR}}}}} \right|}^2}{{\left| {{h_{{\rm{SD}}}}} \right|}^2}/\left({W\sigma ^2}\right)}}{\bf{G}}{{\bf{G}}^{\cal H}}} \right)} \right]}_{{R_{{\rm{D2}}}}}.
\end{align}
\hrulefill
\end{figure*}
The existence of linear mapping matrix $\bf{G}$ in ${R_{{\rm{D2}}}}$ makes ${R_{{\rm{D}}}}$ complex. To make it traceable, we introduce eigen-decomposition, i.e.,  ${\bf{G}}{{\bf{G}}^{\cal H}} \!\!= \!\! {{\bf{U}}^{\cal H}}\!{\bf{\Lambda U}}$, under the power constraint $\frac{1}{N}{\rm{Tr}}\left( {{\bf{G}}{{\bf{G}}^{\cal H}}} \right)\!\! =\!\! 1$ to ensure that the forwarded signals have unit average power, where $\bf{U}$ is a unitary matrix, ${\bf{\Lambda }} = {\rm{diag}}\left( {{\lambda _1}, \cdots ,{\lambda _N}} \right)$ and the eigenvalues, i.e., ${{\lambda _1}, \cdots ,{\lambda _N}}$, are non-negative. Due to the same number and multiplicity of non zero eigenvalues of ${\bf{G}}{{\bf{G}}^{\cal H}}$ and ${{\bf{G}}^{\cal H}}{\bf{G}}$, there are at least $N - \min \left( {M,N} \right)$ zero eigenvalues of ${\bf{G}}{{\bf{G}}^{\cal H}}$.
Accordingly, we assume that the remaining non-negative eigenvalues of ${\bf{G}}{{\bf{G}}^{\cal H}}$, except for $N - \min \left( {M,N} \right)$ zero eigenvalues, are denoted as ${\lambda _1}, \cdots ,{\lambda _{\min \left( {M,N} \right)}}$. As such, ${R_{{\rm{D2}}}}$ is converted into
\begin{align}
{R_{{\rm{D2}}}} \!\!=\!\! \frac{{{T_s}\!W}}{L}\!\sum\limits_{i = 1}^{\min \left( \! {M,N} \! \right)}\!\!\!\!\!\!\!{{{\log }_2}} \!\!\left(\!\! {1 \!\!+\!\! \frac{{{{{\lambda _i}{P_1}{{\left| {{h_{{\rm{RD}}}}} \right|}^2}} \! \! \mathord{\left/
 {\vphantom {{{P_1}{{\left| {{h_{{\rm{RD}}}}} \right|}^2}}{\!\!\left(\!\! {W\!\!{\sigma ^2}} \right)}}} \right.
 \kern-\nulldelimiterspace} {\left( {W{\sigma ^2}} \right)}}}}{{1 \!\!+\!\! {{\beta \!{P_0}{{\left| {{h_{{\rm{SR}}}}} \right|}^2}{{\left| {{h_{{\rm{SD}}}}} \right|}^2}} \mathord{\!\left/\!
 {\vphantom {{\beta {P_0}{{\left| {{h_{{\rm{SR}}}}} \right|}^2}{{\left| {{h_{{\rm{SD}}}}} \right|}^2}} {\left( {W\!\!\!{\sigma ^2}} \right)}}} \right.
 \kern-\nulldelimiterspace}\!{\left( {W{\sigma ^2}} \right)}}}}}\!\!\right)\!\!.
\end{align}

Through the above analysis, the achievable throughput of the IoT node is given as \cite{7274646}
\begin{align}\label{17}
{R_{{\rm{sum}}}} {\rm{      }} = \max \left\{ {{R_{{\rm{SD}}}},\min \left( {{R_{{\rm{SR}}}},{R_{\rm{D}}}} \right)} \right\}.
\end{align}

\emph{Remark 1.} When the time duration for the backscatter and relay phases is unequal, i.e., $M\!\! \ne \!\!N$, the maximum ratio combining cannot be used at the receiver. Accordingly, the existing works \cite{8417655,8490651,9473528} used an upper bound expression of that derived in \cite{7274646}, as discussed in \emph{Remark 2},
leading to overestimate the performance. In this paper, we combine the received signal with linear mapping method, as shown in \eqref{11}, through which the original information can be jointly detected and the system-throughput expression at the receiver can be calculated as \eqref{13}
. Thus, we can obtain the total throughput of the relay-enabled BC system in \eqref{17}. Note that under the equal time allocation, our derived expression equals that of \cite{7274646} and this can be referred to \emph{Remark 2}.

\section{Problem Formulation and Iterative Algorithm}
In this section, we formulate a mixed-integral non-convex problem to maximize the throughput of considered relay-enabled BC network, and then solve it by proposing a low complexity iterative algorithm.

\subsection{Problem Formulation}
Our goal is achieving the maximum throughput by optimizing the linear mapping matrix, the power allocation of the HAP, and the PRC of the IoT node. Accordingly, the optimization problem can be formulated as
\begin{align}\notag
{\mathcal{P}_1}\;\;&\mathop {\max }\limits_{{\lambda _1}, \cdots ,{\lambda _{\min \left( {M,N} \right)}},{P_{\rm{0}}},{P_1},\beta ,M,N} {\rm{ }}\;\;\;\;{R_{{\rm{sum}}}}\\\notag
{\rm{s.t.}}\;\;\;\;&{\rm{  C1 }}:\frac{M}{L}{T_s}{P_{\rm{0}}} + \frac{N}{L}{T_s}{P_1} \le E,\\\notag
&{\rm{                 C2 }}:{E_{\rm{h}}} \ge \frac{M}{L}{T_s}{P_{\rm{c}}},\\\notag
&{\rm{                 C3 }}:M + N = L,\\\notag
&{\rm{                 C4 }}:M,N \ge 0,  \;\;\;\;\forall M,N \in \cal {L},\\\notag
&{\rm{                 C5 }}:\!\!\sum\limits_{i = 1}^{\min \left( {M,N} \right)}\!\!\!\! {{\lambda _i}}=N,\\\notag
&{\rm{                 C6 }}:0 \le {P_{\rm{0}}},{P_1} \le {P_{\max }},\\\label{18}
&{\rm{                 C7 }}:0 \le \beta  \le 1,
\end{align}
where $E$ and $P_{\rm{max}}$ denote the total energy in one transmission block and the maximum transmit power of the HAP, respectively, $\mathcal L= \left\{ {1,2, \ldots ,L} \right\}$, $\mathrm{C}5$ is derived from the power constraint, i.e., $\frac{1}{N}{\rm{Tr}}\left( {{\bf{G}}{{\bf{G}}^{\cal H}}} \right) = 1$, and $P_{\rm{c}}$ is the constant circuit power consumption for BC.

 In ${\mathcal{P}_1}$, $\mathrm{C}1$ denotes the energy constraint of the HAP. $\mathrm{C}2$ ensures the energy-causality constraint of the IoT node, i.e., the consumed energy of the IoT node should not exceed its harvested energy. $\mathrm{C}3$ and $\mathrm{C}4$ constrain the number of subframes  for two transmission phases. $\mathrm{C}6$ and $\mathrm{C}7$ specify the value ranges of the transmit power of the HAP in two phase and the PRC of the IoT node.

Problem ${\mathcal{P}_1}$ is a mixed-integer non-convex optimization problem due to the discrete time variables and the coupling between optimization variables, e.g., $M$ and $P_0$. Thus, this problem cannot be directly solved by using the existing convex tools. In what follows, we solve ${\mathcal{P}_1}$ via a three-step approach.
\subsection{Problem Transformation and Iterative Algorithm}
In the \emph{first step}, we derive the optimal eigenvalues of the linear mapping matrix, i.e., ${{\lambda _1}, \cdots ,{{\lambda _{\min \left( {M,N} \right)}}}}$, via introducing auxiliary variables and exploiting KKT conditions, and substitute them into ${\mathcal{P}_1}$ for reducing optimization variables.

It can be observed that the max-min expression of the objective function brings challenges to solve ${\mathcal{P  }_1}$. To address this, we first introduce an auxiliary variable $t = \min \left\{ {{R_{{\rm{SR}}}},{{R_{\rm{D}}}}} \right\}$. Then, ${\mathcal{P  }_1}$ can be reformulated as
\begin{align}\notag
{\mathcal{P}_2}\;\;&\mathop {\max }\limits_{{\lambda _1}, \cdots ,{\lambda _{\min \left( {M,N} \right)}},{P_{\rm{0}}},{P_1},\beta ,M,N,t} {\rm{ }}\max \left\{ {{{R_{{\rm{SD}}}}},t} \right\}\\\notag
{\rm{s.t.}}\;\;\;\;&{\rm{   C1}} - {\rm{C7}},\\\notag
&{\rm{C8}}:{{R_{{\rm{SR}}}}} \ge t,\\\label{20}
&{\rm{C9}}:{{R_{\rm{D}}}} \ge t.
\end{align}

\textbf{Lemma 1.}
To maximize ${\mathcal{P}_2}$, the equality of ${\rm{C8}}$ or ${\rm{C9}}$ must be satisfied, then an equivalent transformation between ${\mathcal{P}_1}$ and ${\mathcal{P}_2}$ is completed.

\emph{Proof:} Please see Appendix  A. \hfill {$\blacksquare $}

${\mathcal{P  }_2}$ is still non-convex. In what follows, by leveraging the KKT conditions, we can obtain some useful results, as summarized in \textbf{Lemma2}.

\textbf{Lemma 2.}
To maximize the throughput in ${\mathcal{P  }_2}$, the remaining eigenvalues, except for zero eigenvalues, should be equal, i.e., $\lambda _1^* =  \cdots  = \lambda _{\min \left( {M,N} \right)}^* = \frac{N}{{\min \left( {M,N} \right)}}$, where $*$ denotes the optimal solution corresponding to the optimization variables.

\emph{Proof:} Please see Appendix B. \hfill {$\blacksquare $}

\textbf{Theorem 1:} Based on \textbf{Lemma 2}, the optimal linear mapping matrix $\bf{G}^*$ to maximize the throughput should satisfy ${\bf{G}^*}{({\bf{G}^*})^{\mathcal{H}}} = {{\bf{I}}_N}$ under $M \ge N$, while ${({\bf{G}^*})^{\mathcal{H}}}{\bf{G}^*} = {\frac{N}{M}}{{\bf{I}}_M}$ under $M<N$.

By substituting all eigenvalues of ${\bf{G}^*}{({\bf{G}^*})^{\mathcal{H}}}$ into ${\mathcal{P}_2}$, we obtain an equivalent optimization problem as
\begin{align}\notag
{\mathcal{P}_3}\;\;&\mathop {\max }\limits_{{P_{\rm{0}}},{P_1},\beta ,M,N,t} {\rm{ }}\max \left\{ {{{R_{{\rm{SD}}}}},t} \right\}\\\notag
{\rm{s.t.}}\;\;\;\;&{\rm{   C1}} - {\rm{C4}},{\rm{   C6}}-{\rm{   C8}},\\\label{21}
&{\rm{C9}}\!-\!1:{{R'_{\rm{D}}}} \ge t,
\end{align}
where ${{R'_{\rm{D}}}}$ is obtained by substituting all eigenvalues into ${{R_{\rm{D}}}}$, i.e., ${{R'_{\rm{D}}}} = {R_{{\rm{SD}}}} + {{R'_{{\rm{D2}}}}}$,
${R'_{{\rm{D2}}}} = \frac{{{T_s}W}}{L}\sum\limits_{i = 1}^{\min \left( {M,N} \right)}{{{\log }_2}} $
$\left( {1 + \frac{{{{{P_1}{{\left| {{h_{{\rm{RD}}}}} \right|}^2}} \mathord{\left/
 {\vphantom {{{P_1}{{\left| {{h_{{\rm{RD}}}}} \right|}^2}} {\left( {W{\sigma ^2}} \right)}}} \right.
 \kern-\nulldelimiterspace} {\left( {W{\sigma ^2}} \right)}}}}{{1 + {{\beta {P_0}{{\left| {{h_{{\rm{SR}}}}} \right|}^2}{{\left| {{h_{{\rm{SD}}}}} \right|}^2}} \mathord{\left/
 {\vphantom {{\beta {P_0}{{\left| {{h_{{\rm{SR}}}}} \right|}^2}{{\left| {{h_{{\rm{SD}}}}} \right|}^2}} {\left( {W{\sigma ^2}} \right)}}} \right.
 \kern-\nulldelimiterspace} {\left( {W{\sigma ^2}} \right)}}}} \times \frac{N}{{\min \left( {M,N} \right)}}} \right).$  The specific expression of ${{R'_{\rm{D}}}}$ is shown at the top of the next page.
\begin{figure*}
\begin{align}
{{R'_{\rm{D}}}} = \left\{ {\begin{array}{*{20}{c}}
{\frac{{{T_s}WN}}{L}{{\log }_2}\left( {1 + \frac{{\beta {P_0}{{\left| {{h_{{\rm{SR}}}}} \right|}^2}{{\left| {{h_{{\rm{SD}}}}} \right|}^2} + {P_1}{{\left| {{h_{{\rm{RD}}}}} \right|}^2}}}{{W{\sigma ^2}}}} \right) + \frac{{{T_s}W\left( {M - N} \right)}}{L}{{\log }_2}\left( {1 + \frac{{\beta {P_0}{{\left| {{h_{{\rm{SR}}}}} \right|}^2}{{\left| {{h_{{\rm{SD}}}}} \right|}^2}}}{{W{\sigma ^2}}}} \right),{\rm{   }}M \ge N},\\
\!\!\!\!\!\!\!\!\!\!\!\!\!\!\!\!\!\!\!\!\!\!\!\!
\!\!\!\!\!\!\!\!\!\!\!\!\!\!\!\!\!\!\!\!\!\!\!\!
\!\!\!\!\!\!\!\!\!\!\!\!\!\!\!\!\!\!\!\!\!\!\!\!
\!\!\!\!\!\!\!\!\!\!\!\!\!\!\!\!\!
{\frac{{{T_s}WM}}{L}{{\log }_2}\left( {1 + \frac{{\beta {P_0}{{\left| {{h_{{\rm{SR}}}}} \right|}^2}{{\left| {{h_{{\rm{SD}}}}} \right|}^2}}}{{W{\sigma ^2}}} + \frac{{N{P_1}{{\left| {{h_{{\rm{RD}}}}} \right|}^2}}}{{MW{\sigma ^2}}}} \right),{\rm{   }}M < N{\rm{ }}}.
\end{array}} \right.
\end{align}
\hrulefill
\end{figure*}

\emph{Remark 2:}
It is worth noting that the system-throughput expression employed in the previous studies, e.g.,  \cite{8417655,8490651,9473528}, is an upper bound of that we derived in this work, and the reasons are as follows. In \cite{8417655,8490651,9473528}, the system-throughput expression\footnote{In \cite{8417655,8490651,9473528}, the authors assumed that the channel gain of the direct transmission is much smaller than that of the source-relay link, and thus ignored the comparison between ${R_{{\rm{SD}}}}$ and ${R_{{\rm{SR}}}}$. Since this work has not made assumption on the channel gains of all links, for traceability analysis, we made such a comparison.} is calculated as ${R_{{\rm{sum0}}}} = \max \left\{ {{R_{{\rm{SD}}}},\min \left( {{R_{{\rm{SR}}}},{R_{{\rm{D0}}}}} \right)} \right\}$, where ${R_{{\rm{D0}}}} = {R_{{\rm{SD}}}} + {R_{{\rm{RD}}}}$ and ${R_{{\rm{RD}}}} = \frac{{{T_s}WN}}{L}{\log _2}\left( {1 + \frac{{{P_1}{\left| {{h_{{\rm{RD}}}}} \right|^2}}}{{W{\sigma ^2}}}} \right)$, while in our work, it is expressed as ${R_{{\rm{sum}}}} = \max \left\{ {{R_{{\rm{SD}}}},\min \left( {{R_{{\rm{SR}}}},{R'_{\rm{D}}}} \right)} \right\}$, where ${{R'_{\rm{D}}}} = {R_{{\rm{SD}}}} + {{R'_{{\rm{D2}}}}}$. The reason why ${R_{{\rm{sum0}}}}$ is an upper bound of  ${R_{{\rm{sum}}}}$ lies in ${R_{{\rm{RD}}}} \ge {{R'_{{\rm{D2}}}}}$, which can be interpreted from three cases in \eqref{100}, as shown at the top of the next page. In particular, under the equal time allocation, the derived system-throughput expression in this work can be reduced to ${R_{{\rm{sum}}}}\!\!= \!\! \frac{{{T_s}W}}{2}{\log _2}\left( {1 + \max [{\gamma _{{\rm{SD}}}},\min [{\gamma _{{\rm{SR}}}},{\gamma _{{\rm{SD}}}} + {\gamma _{{\rm{RD}}}}]]} \right)$, which is the same as the expression of system throughput in \cite{7274646}, where ${{\gamma _{{\rm{SD}}}}}$, ${{\gamma _{{\rm{SR}}}}}$ and ${{\gamma _{{\rm{RD}}}}}$ denote the signal to interference plus noise ratio (SINR) of the link of S$-$D, S$-$R and R$-$D, i.e., ${\gamma _{{\rm{SD}}}} \!\!=\! \! \frac{{\beta {P_0}{{\left| {{h_{{\rm{SR}}}}} \right|^2}{\left| {{h_{{\rm{SD}}}}} \right|^2}} }}{{W{\sigma ^2}}}$, ${\gamma _{{\rm{SR}}}}\!\!= \!\!\frac{{\beta {P_0}{\left| {{h_{{\rm{SR}}}}} \right|^4}}}{{W{\sigma ^2}}}$, and ${\gamma _{{\rm{RD}}}}\!\!= \!\! \frac{{{P_1}{\left| {{h_{{\rm{RD}}}}} \right|^2}}}{{W{\sigma ^2}}}$. While in the previous studies \cite{8417655,8490651,9473528},  ${{R_{{\rm{sum0}}}}}$, under the equal time allocation, equals $ \frac{{{T_s}W}}{2}{\log _2}\!\left( {1\! \!+ \!\max [{\gamma _{{\rm{SD}}}},\!\min [{\gamma _{{\rm{SR}}}},\!{\gamma _{{\rm{SD}}}}\!\! + \! {\gamma _{{\rm{RD}}}}\!\!+\!{\gamma _{{\rm{SD}}}}\!{\gamma _{{\rm{RD}}}}]]} \right)$, which is an upper bound of the accurate system throughput evidently. This not only shows the accuracy of the expression obtained in this paper, but also proves that the expression adopted in \cite{8417655,8490651,9473528} is an upper bound of what we derived.
\begin{figure*}
\begin{align}\label{100}
\left\{
  \begin{array}{ll}
    \rm{Case}(i), & \hbox{${R_{{\rm{SR}}}} \ge {R_{{\rm{D0}}}} \Rightarrow { \left\{
  \begin{array}{ll}
  {R_{{\rm{sum0}}}} ={R_{{\rm{D0}}}} \\
  {R_{{\rm{sum}}}}={R'_{\rm{D}}}
  \end{array}
   \right.} \Rightarrow{R_{{\rm{sum0}}}} \ge {R_{{\rm{sum}}}}$;} \\
    \rm{Case}(ii), & \hbox{$ {R_{{\rm{SR}}}}< {{R'_{\rm{D}}}} \Rightarrow
\left\{
  \begin{array}{ll}
{R_{{\rm{sum0}}}} = \max \left\{ {{R_{{\rm{SD}}}},{R_{{\rm{SR}}}}} \right\} \\
{R_{{\rm{sum}}}} = \max \left\{ {{R_{{\rm{SD}}}},{R_{{\rm{SR}}}}} \right\}
  \end{array}
\right.
\Rightarrow {R_{{\rm{sum0}}}}={R_{{\rm{sum}}}}$;} \\
    \rm{Case}(iii), & \hbox{${{R'_{\rm{D}}}}\le {R_{{\rm{SR}}}}< {R_{{\rm{D0}}}}\Rightarrow
\left\{
  \begin{array}{ll}
    {R_{{\rm{sum0}}}}= {{R_{{\rm{SR}}}}} \\
{{R_{{\rm{sum}}}}={{R'_{\rm{D}}}}}
  \end{array}
\right.
\Rightarrow {R_{{\rm{sum0}}}}\ge {R_{{\rm{sum}}}}$.}
  \end{array}
\right.
\end{align}
\hrulefill
\end{figure*}

${\mathcal{P  }_3}$ is still a non-convex mixed-integer programming problem although it has been simplified. In the \emph{second step}, we transform ${\mathcal{P  }_3}$ into a continuous variable optimization problem by applying the time-sharing relaxation which is proposed in \cite{793310}, and then solve it from two cases, as shown in ${\mathcal{P  }_{5.1}}$ and ${\mathcal{P  }_{5.2}}$, respectively. In specific, a continuous auxiliary real variable $\rho$ in the closed interval $[0,1]$ is introduced to relax discrete variable\footnote{The gap caused by the time-sharing method between the original mixed-integer optimization problem, i.e., ${\mathcal{P  }_3}$, and the converted continuous optimization problem, i.e., ${\mathcal{P  }_4}$, decreases with the increase of the number of total subframes, i.e., $L$, which can be verified in Fig. 4 of simulation.} $\frac{M}{L}$, where $\rho$ can be considered as the time allocation factor. Then, problem ${\mathcal{P  }_3}$ can be converted into
\begin{align}\notag
{\mathcal{P}_4}\;\;&\mathop {\max }\limits_{{P_{\rm{0}}},{P_1},\beta ,\rho ,t} {\rm{ }}\max \left\{ {{{R'_{{\rm{SD}}}}},t} \right\}\\\notag
{\rm{s.t.}}\;\;\;\;&{\rm{C1}} \!-\! 1:\rho {P_{\rm{0}}} + \left( {1 - \rho } \right){P_1} \le P,\\\notag
&{\rm{C2}}\!-\!1:\eta \left( {1 - \beta } \right){P_0}{\left| {{h_{{\rm{SR}}}}} \right|^2} \ge {P_{\rm{c}}},\\\notag
&{\rm{C6}}- {\rm{C7}},\\\notag
&{\rm{C8}}\!-\!1:{{R'_{{\rm{SR}}}}} \ge t,\\\notag
&{\rm{C9}}\!-\!2:{{R''_{\rm{D}}}} \ge t,\\
&{\rm{C10}}:0\le\rho \le 1,\label{24}
\end{align}
where $P \!\!=\!\! E/{T_s}$,
${{R'_{{\rm{SD}}}}} \!\!=\! \!{T_s}W\!\rho {\log _2}\!\left(\! {1\! +\! \frac{{\beta {P_0}{\left| {{h_{{\rm{SR}}}}} \right|^2}{\left| {{h_{{\rm{SD}}}}} \right|^2}}}{{W{\sigma ^2}}}} \!\right)$,
${{R'_{{\rm{SR}}}}}\! =\!{T_s}W\!\rho {\log _2}\!\left(\! {1\!+\!\frac{{\beta {P_0}{\left| {{h_{{\rm{SR}}}}} \right|^4}}}{{W{\sigma ^2}}}}\! \right)$.
${{R''_{\rm{D}}}}$ is shown as the top of the next page.
\begin{figure*}
\begin{align}
{{R''_{\rm{D}}}} = \left\{ {\begin{array}{*{20}{c}}
{{T_s}W\left( {1 - \rho } \right){{\log }_2}\left( {1 + \frac{{\beta {P_0}{{\left| {{h_{{\rm{SR}}}}} \right|}^2}{{\left| {{h_{{\rm{SD}}}}} \right|}^2} + {P_1}{{\left| {{h_{{\rm{RD}}}}} \right|}^2}}}{{W{\sigma ^2}}}} \right) + {T_s}W\left( {2\rho  - 1} \right){{\log }_2}\left( {1 + \frac{{\beta {P_0}{{\left| {{h_{{\rm{SR}}}}} \right|}^2}{{\left| {{h_{{\rm{SD}}}}} \right|}^2}}}{{W{\sigma ^2}}}} \right),{\rm{   }}\frac{1}{2} \le \rho  \le 1},\\
\!\!\!\!\!\!\!\!\!\!\!\!\!\!\!\!\!\!\!\!\!\!\!\!
\!\!\!\!\!\!\!\!\!\!\!\!\!\!\!\!\!\!\!\!\!\!\!\!
\!\!\!\!\!\!\!\!\!\!\!\!\!\!\!\!\!\!\!\!\!\!\!\!
\!\!\!\!\!\!\!\!\!\!\!\!\!\!\!\!\!\!\!\!\!\!\!\!
\!\!\!\!\!\!\!\!
{{T_s}W\rho {{\log }_2}\left( {1 + \frac{{\beta {P_0}{{\left| {{h_{{\rm{SR}}}}} \right|}^2}{{\left| {{h_{{\rm{SD}}}}} \right|}^2}}}{{W{\sigma ^2}}} + \frac{{\left( {1 - \rho } \right){P_1}{{\left| {{h_{{\rm{RD}}}}} \right|}^2}}}{{\rho W{\sigma ^2}}}} \right),{\rm{   }}0 \le \rho < \frac{1}{2}{\rm{                                                           }}}.
\end{array}} \right.
\end{align}
\hrulefill
\end{figure*}
Next, we need to explore the optimal solutions of ${\mathcal{P  }_4}$. Towards this end, \textbf{Lemma 3} is provided as follows.

\textbf{Lemma 3.}
For the optimal solutions, the equalities of constraints C1$-$1 and C2$-$1 must hold. Therefore, the optimal PRC, ${\beta ^ * }$, is $1 - \frac{{{P_{\rm{c}}}}}{{\eta {P_{\rm{0}}}{\left| {{h_{{\rm{SR}}}}} \right|^2}}}$. This means the total energy of the HAP and the harvested energy of the IoT node should be used up for maximizing the throughput.

\emph{Proof:} Please see Appendix C. \hfill {$\blacksquare $}

Based on \textbf{Lemma 3}, we substitute the optimal PRC, ${\beta ^ * }$, into ${\mathcal{P  }_4}$, which results in
\begin{align}\notag
{\mathcal{P}_5}\;\;&\mathop {\max }\limits_{{P_{\rm{0}}},{P_1},\rho ,t} {\rm{ }}\max \left\{ {{{R''_{{\rm{SD}}}}},t} \right\}\\\notag
{\rm{s.t.}}\;\;\;\;&{\rm{C1}}\! - \!1:\rho {P_{\rm{0}}} + \left( {1 - \rho } \right){P_1} \le P{\rm{,}}\\\notag
&{\rm{C6}},{\rm{C10}},\\\notag
&{\rm{C7}} \!- \!1:{P_{\rm{c}}} \le \eta {P_{\rm{0}}}{\left| {{h_{{\rm{SR}}}}} \right|^2}{\rm{,}}\\\notag
&{\rm{C8}}\! -\! 2:{{R''_{{\rm{SR}}}}} \ge t,\\\label{28}
&{\rm{C9}}\! -\! 3:{{R'''_{\rm{D}}}} \ge t,
\end{align}
where ${{R''_{{\rm{SD}}}}}{\rm{= }}{T_s}W\rho {\log _2}\left({B+\frac{{{P_0}{\left| {{h_{{\rm{SR}}}}}\right|^2}{\left| {{h_{{\rm{SD}}}}}\right|^2}}}{{W {\sigma ^2}}}} \right){\rm{ }}$,${{R''_{{\rm{SR}}}}}={T_s}W\rho$
${\log _2}\left( {A+ \frac{{{P_0}{\left| {{h_{{\rm{SR}}}}} \right|^4}}}{{W{\sigma ^2}}}}\right)$,
$A = 1 - \frac{{{P_{\rm{c}}}{\left| {{h_{{\rm{SR}}}}} \right|^2}}}{{\eta W{\sigma ^2}}}$, $B = 1 - \frac{{{P_{\rm{c}}}{\left| {{h_{{\rm{SD}}}}} \right|^2}}}{{\eta W{\sigma ^2}}}$.
${{R'''_{\rm{D}}}}$ is shown as the top of the next page.
\begin{figure*}
\begin{align}
{{R'''_{\rm{D}}}} = \left\{ {\begin{array}{*{20}{c}}
{{T_s}W\left( {1 - \rho } \right){{\log }_2}\left( {B + \frac{{{P_0}{{\left| {{h_{{\rm{SR}}}}} \right|}^2}{{\left| {{h_{{\rm{SD}}}}} \right|}^2} + {P_1}{{\left| {{h_{{\rm{RD}}}}} \right|}^2}}}{{W{\sigma ^2}}}} \right) + {T_s}W\left( {2\rho  - 1} \right){{\log }_2}\left( {B + \frac{{{P_0}{{\left| {{h_{{\rm{SR}}}}} \right|}^2}{{\left| {{h_{{\rm{SD}}}}} \right|}^2}}}{{W{\sigma ^2}}}} \right),{\rm{   }}\frac{1}{2} \le \rho  \le 1},\\
\!\!\!\!\!\!\!\!\!\!\!\!\!\!\!\!\!\!\!\!\!\!\!\!
\!\!\!\!\!\!\!\!\!\!\!\!\!\!\!\!\!\!\!\!\!\!\!\!
\!\!\!\!\!\!\!\!\!\!\!\!\!\!\!\!\!\!\!\!\!\!\!\!
\!\!\!\!\!\!\!\!\!\!\!\!\!\!\!\!\!\!\!\!\!\!\!\!
\!\!\!\!\!\!\!\!
{{T_s}W\rho {{\log }_2}\left( {B + \frac{{{P_0}{{\left| {{h_{{\rm{SR}}}}} \right|}^2}{{\left| {{h_{{\rm{SD}}}}} \right|}^2}}}{{W{\sigma ^2}}} + \frac{{\left( {1 - \rho } \right){P_1}{{\left| {{h_{{\rm{RD}}}}} \right|}^2}}}{{\rho W{\sigma ^2}}}} \right),{\rm{   }}0 \le \rho  < \frac{1}{2}{\rm{                                                           }}}.
\end{array}} \right.
\end{align}
\hrulefill
\end{figure*}

It can be seen from \eqref{28} that the objective function includes a max function and it makes this problem non-convex. Thus, we discuss ${\mathcal{P  }_5}$ from the following two cases so that the max function can be removed.

\emph{Case 1 with ${R''_{{\rm{SD}}}} \le t$:} In this case, we have ${R''_{{\rm{SD}}}} \le {R''_{{\rm{SR}}}}$ and ${R''_{{\rm{SD}}}} \le {R'''_{\rm{D}}}$, which infers ${h_{{\rm{SD}}}} \le {h_{{\rm{SR}}}}$. Accordingly, ${\mathcal{P  }_5}$ can be written as
\begin{align}\notag
{\mathcal{P}_{5.1}}\;\;&\mathop {\max }\limits_{{P_{\rm{0}}},{P_1},\rho ,t} {\rm{ }}t\\\label{33}
{\rm{s.t.}}\;\;\;\;&{\rm{C1}} \!-\! 1,{\rm{ C6,C7}} \!-\! 1,{\rm{C8}}\!-\!2,{\rm{ C9}}\! - \!3,{\rm{ C10}} .
\end{align}

For analysis, ${\mathcal{P}_{5.1}}$ can be discussed from the following scenarios, i.e., ${1 \mathord{\left/
 {\vphantom {1 2}} \right.
 \kern-\nulldelimiterspace} 2} \le \rho  \le 1$ and $0 \le \rho  < {1 \mathord{\left/
 {\vphantom {1 2}} \right.
 \kern-\nulldelimiterspace} 2}$.

When ${1 \mathord{\left/
 {\vphantom {1 2}} \right.
 \kern-\nulldelimiterspace} 2} \le \rho  \le 1$, ${\mathcal{P  }_{5.1}}$ can be expressed as ${\mathcal{P  }_{5.1.1}}$,
\begin{align}\notag
{{\cal P}_{5.1.1}}\;\;&{\rm{     }}\mathop {\max }\limits_{{P_{\rm{0}}},{P_1},\rho ,t} t\\\notag
{\rm{s.t.}}\;\;\;\;&{\rm{C1}}\!-\!1,{\rm{C6}},{\rm{C7}}\!-\!1,{\rm{C8}} \!-\!2,{\rm{C9}}\!-\!3,\\
&{\rm{C10}}\!-\!1:{1 \mathord{\left/
 {\vphantom {1 2}} \right.
 \kern-\nulldelimiterspace} 2} \le \rho  \le 1.
\end{align}
It can be observed that the objective function is linear and easy to handle, while the auxiliary variable $\rho$ is still coupled with ${P_0}$ and ${P_1}$ in C1$-$1, C8$-$2 and C9$-$3. To decouple it, we construct two auxiliary variables, i.e., $a = {P_0}\left( {1 - \rho } \right)$ and $b = {P_1}\left( {1 -\rho } \right)$, and substitute ${P_0} = \frac{a}{{1-\rho}}$ and ${P_1} = \frac{b}{{1 - \rho }}$ into ${P_0}$ and ${P_1}$, resulting in
\begin{align}\notag
{{{\cal P}_{5.1.2}}\;\;}&{\mathop {\max }\limits_{a,b,\rho ,t} t}\\\notag
{{\rm{s}}.{\rm{t}}.\;\;\;\;}&{{\rm{C1}}\!-\!{\rm{2}}:a - b + \frac{b}{\rho } \le P\left( {\frac{1}{\rho } - 1} \right),}\\\notag
{}&{{\rm{C6}}\!-\!{\rm{1}}:0 \le a,b \le {P_{\max }}(1 - \rho ),}\\\notag
{}&{{\rm{C7}} \!-\! 2:{P_{\rm{c}}}(1 - \rho ) \le \eta a{{\left| {{h_{{\rm{SR}}}}} \right|}^2},}\\\notag
{}&{{\rm{C8}}\!-\!3:{{R'''_{{\rm{SR}}}} }\ge t,}\\\notag
{}&{{\rm{C9}}\!-\!{\rm{4}}:{{\tilde R_{\rm{D}}} }\ge t,}\\
{}&{{\rm{C10}}\!-\!1,}
\end{align}
where ${{\tilde R}_{\rm{D}}} \!\!=\!\! {T_s}\!W\!\left( {1 \!\!-\!\! \rho } \right)\!{\log _2}\!\left( \!{B \!+ \!\frac{{a{\left| {{h_{{\rm{SR}}}}} \right|^2}{\left| {{h_{{\rm{SD}}}}} \right|^2} + b{\left| {{h_{{\rm{RD}}}}} \right|^2}}}{{\left( {1 - \rho } \right)W{\sigma ^2}}}}\! \right)+ {T_s}W$
$\left({2\rho \! -\!\! 1} \!\right)\!{\log _2}\!\!\left(\!{B \!\!+\!\! \frac{{a{\left| {{h_{{\rm{SR}}}}}\right|^2}{\left| {{h_{{\rm{SD}}}}}\right|^2}}}{{\left( {1-\rho } \right)W{\sigma ^2}}}} \!\right)$,
${{R'''_{{\rm{SR}}}}} \!\!=\!\! {T_s}\!W\!\rho {\log _2}\!\!\left(\!{A\!\! +\!\! \frac{{a{\left| {{h_{{\rm{SR}}}}} \right|^4}}}{{\left(\! {1- \rho }\! \right)W{\sigma ^2}}}} \!\right)$.

To tackle these non-convex constraints, i.e., C1$-$2, C8$-$3, and C9$-$4, SCA technology \cite{8675438} is employed to obtain a locally optimal solution in an iterative manner that is shown in \textbf{Lemma 4}.

\textbf{Lemma 4.}
For any given ${\rho ^j}$, ${a^j}$ and ${b^j}$, we have
\begin{align}
y\left(\rho  \right) \ge {y_{lb}}\left( {{\rho ^j}} \right), f\left( {\rho ,b} \right) \le {f_{ub}}\left( {{\rho ^j},{b^j}} \right),
\end{align}
\begin{align}
g\left( {\rho ,a} \right) \ge {g_{lb}}\left( {{\rho ^j},{a^j}} \right), w\left( {\rho ,a} \right) \ge {w_{lb}}\left( {{\rho ^j},{a^j}} \right),
\end{align}
where the above equalities hold at the points $\rho  = {\rho ^j}$, $a = {a^j}$ and $b = {b^j}$, and the functions of $y\left( \rho  \right)$, $f\left( {\rho ,b} \right)$, $g\left( {\rho ,a} \right)$, $w\left( {\rho ,a} \right)$, $e\left( {\rho ,a, b} \right)$,
 ${y_{lb}}\left( {{\rho ^j}} \right)$, ${f_{ub}}\left( {{\rho ^j},{b^j}} \right)$, ${g_{lb}}\left( {{\rho ^j},{a^j}} \right)$ and ${w_{lb}}\left( {{\rho ^j},{a^j}} \right)$ are defined in Appendix D.
At the local points ${\rho ^j}$, ${{a^j}}$ and ${{b^j}}$, both the functions $y\left( \rho  \right)$, $f\left( {\rho ,b} \right)$, $g\left( {\rho ,a} \right)$, $w\left( {\rho ,a} \right)$ and its corresponding bounds
${y_{lb}}\left( {{\rho ^j}} \right)$,
${f_{ub}}\left( {{\rho ^j},{b^j}} \right)$, ${g_{lb}}\left( {{\rho ^j},{a^j}} \right)$, ${w_{lb}}\left( {{\rho ^j},{a^j}} \right)$ have the identical gradients.

\emph{Proof:} Please refer to Appendix D. \hfill {$\blacksquare $}

By performing SCA, ${\mathcal{P  }_{5.1.2}}$ can be transformed to a convex one, which is given by
\begin{align}\notag
{\mathcal{P}_{5.1.3}}\;\;&\mathop {\max }\limits_{a,b,\rho ,t} {\rm{ }}t\\\notag
{\rm{s.t.}}\;\;\;\;&{\rm{C1}} \!- \! {\rm{3}}:a - b + {f_{ub}}\left( {{\rho ^j},{b^j}} \right) \le P\left( {{y_{lb}}\left( {{\rho ^j}} \right) - 1} \right),\\\notag
&{\rm{C6}} \!- \!{\rm{1,C7}} \!- \! 2,{\rm{C}}10 \!- \!1,\\\notag
&{\rm{C8}} \!-\! {\rm{4}}:g_{lb}\left( {\rho^j ,a^j} \right) \ge t,{\rm{ }}\\\label{40}
&{\rm{C9}} \!- \!5:e\left( {\rho ,a, b} \right) + w_{lb}\left( {\rho^j ,a^j} \right) \ge t.
\end{align}
It can be ready proven that ${\mathcal{P  }_{5.1.3}}$ is convex and thus it can be directly solved by CVX \cite{8675437} based on SCA technology, as summarized in Algorithm 1.

\begin{algorithm}[!t]
\setstretch{1}
    \caption{SCA-based Iteration Algorithm for Solving ${\mathcal{P}_{5.1}}$}
    \label{alg:1}
    \renewcommand{\algorithmicrequire}{\textbf{Input:}}
    \renewcommand{\algorithmicensure}{\textbf{Output:}}
    \begin{algorithmic}[1]
    	\REQUIRE Coordinates of the IoT node, the HAP and the receiver
		\ENSURE ${P_0^*}$, ${P_1^*}$, ${{\rho ^*}}$, ${\beta ^*}$, ${t ^*}$
      \STATE Initialize ${\rho ^0}$, ${a^0}$, ${b^0}$;
        \STATE $\textbf{repeat:}$ \\
        \STATE Obtain $\left\{ {{a^ {*1} },{b^ {*1}},{\rho ^{*1}}, {t^ {*1}} } \right\}$ by solving ${\mathcal{P  }_{5.1.3}}$;
        \STATE Update the obtained continuous optimal solutions ;
        \STATE $\textbf{until:}$ $\left\{ {{a^ {*1} },{b^ {*1}},{\rho ^{*1}}, {t^ {*1}} } \right\}$ is converge;
        \STATE Compute $\left\{ {P_0^{*1},P_1^{*1},{\beta ^{*1}}} \right\}$ by $P_0^{*1}= \frac{{{a^ {*1} }}}{{1 - {\rho ^{*1}}}}$, $P_1^{*1} = \frac{{{b^ {*1} }}}{{1 - {\rho ^{*1}}}}$, ${\beta ^{*1}} = 1 - \frac{{{P_{\rm{c}}}}}{{\eta P_0^{*1}{\left| {{h_{{\rm{SR}}}}} \right|^2}}}$;
\STATE Obtain $\left\{ {u^{*2},v^{*2},{\rho ^{*2}},{t ^{*2}}} \right\}$ by solving ${\mathcal{P}_{5.1.5}}$ ;
        \STATE Compute $\left\{ {P_0^{*2},P_1^{*2},{\beta ^{*2}}} \right\}$ by $P_0^{*2}= \frac{{{u^ {*2} }}}{{{\rho ^{*2}}}}$, $P_1^{*2} = \frac{{{v^ {*2} }}}{{1 - {\rho ^{*2}}}}$, ${\beta ^{*2}} = 1 - \frac{{{P_{\rm{c}}}}}{{\eta P_0^{*2}{\left| {{h_{{\rm{SR}}}}} \right|^2}}}$;
        \IF{${t^{*1}} \ge {t^{*2}}$}
  \STATE $\left\{ {P_0^*,P_1^*,{\rho ^*},{\beta ^*},{t^*}} \right\}=\left\{ {P_0^{*1},P_1^{*1},{\rho ^{*1}},{\beta ^{*1}},{t ^{*1}}} \right\}$;
  \ELSE
  \STATE $\left\{ {P_0^*,P_1^*,{\rho ^*},{\beta ^*},{t^*}} \right\}=\left\{ {P_0^{*2},P_1^{*2},{\rho ^{*2}},{\beta ^{*2}},{t ^{*2}}} \right\}$;
   \ENDIF
    \end{algorithmic}
\end{algorithm}

When ${\rm{   }}0 \le \rho  < {1 \mathord{\left/
 {\vphantom {1 2}} \right.
 \kern-\nulldelimiterspace} 2} $, ${\mathcal{P  }_{5.1}}$ can be expressed as ${\mathcal{P  }_{5.1.4}}$,
\begin{align}\notag
{{\cal P}_{5.1.4}}\;\;&{\rm{     }}\mathop {\max }\limits_{{P_{\rm{0}}},{P_1},\rho ,t} t\\\notag
{\rm{s.t.}}\;\;\;\;&{\rm{C1}}\!-\!1,{\rm{C6}},{\rm{C7}}\!-\!1,{\rm{C8}} \!-\!2,{\rm{C9}}\!-\!3,\\
&{\rm{C10}}\!-\!2:0 \le \rho  < {1 \mathord{\left/
 {\vphantom {1 2}} \right.
 \kern-\nulldelimiterspace} 2}.
\end{align}

Under this scenario, we solve this optimization problem by introducing the following auxiliary variables, i.e., $u = {P_0}\rho $ and $v = {P_1}\left( {1 - \rho } \right)$. Then, ${\mathcal{P}_{5.1.4}}$ is transformed into
\begin{align}\notag
{{P_{5.1.5}}\;\;}&{\mathop {\max }\limits_{u,v,\rho ,t} t}\\\notag
{{\rm{s}}.{\rm{t}}.\;\;\;\;}&{{\rm{C1}}\!-\!{\rm{4}}:u + v \le P,}\\\notag
&{{\rm{C6}}\!-\! {\rm{2}}:0 \le u \le {P_{\max }}\rho ,}\\\notag
&\;\;\;\;\;\;\;\;\;\;\;\;{{\rm{}}0 \le v \le {P_{\max }}(1 - \rho ),}\\\notag
{}&{{\rm{C7}}\!-\! 3:{P_{\rm{c}}}\rho  \le \eta u{{\left| {{h_{{\rm{SR}}}}} \right|}^2},}\\\notag
{}&{{\rm{C8}}\!-\!5:{{\mathord{\buildrel{\lower3pt\hbox{$\scriptscriptstyle\smile$}}
\over R} }_{{\rm{SR}}}} \ge t,}\\\notag
{}&{{\rm{C9}}\!-\!{\rm{6}}:{{\mathord{\buildrel{\lower3pt\hbox{$\scriptscriptstyle\smile$}}
\over R} }_{\rm{D}}} \ge t,}\\
{}&{{\rm{C10}} \!-\! {\rm{2}}:0 \le \rho  < {1 \mathord{\left/
 {\vphantom {1 2}} \right.
 \kern-\nulldelimiterspace} 2},}
\end{align}
where ${{\mathord{\buildrel{\lower3pt\hbox{$\scriptscriptstyle\smile$}}
\over R} }_{\rm{D}}}= {T_s}W\rho {\log _2}\left( {B + \frac{{u{{\left| {{h_{{\rm{SR}}}}} \right|}^2}{{\left| {{h_{{\rm{SD}}}}} \right|}^2}}}{{\rho W{\sigma ^2}}} + \frac{{v{{\left| {{h_{{\rm{RD}}}}} \right|}^2}}}{{\rho W{\sigma ^2}}}} \right)$ and
${{\mathord{\buildrel{\lower3pt\hbox{$\scriptscriptstyle\smile$}}
\over R} }_{{\rm{SR}}}}={T_s}W\rho {\log _2}\left( {A + \frac{{u{{\left| {{h_{{\rm{SR}}}}} \right|}^4}}}{{\rho W{\sigma ^2}}}} \right)$.
Based on the above operations, we obtain an equivalent convex problem, and it can be solved by CVX. After obtaining the approaches to solve
${\mathcal{P  }_{5.1.3}}$ and ${\mathcal{P  }_{5.1.5}}$, problem ${\mathcal{P  }_{5.1}}$ can be solved by designing Algorithm 1.


\emph{Case 2 with ${R''_{{\rm{SD}}}} > t$:} In this case, ${{R''_{{\rm{SD}}}}} > {{R''_{{\rm{SR}}}}}$ or ${{R''_{{\rm{SD}}}}} > {{R'''_{\rm{D}}}}$ should be satisfied. We note that ${{R''_{{\rm{SD}}}}} \le {{R'''_{\rm{D}}}}$ holds. This is because ${{R'''_{\rm{D}}}}$ is the throughput to combine the S$-$D link and the R$-$D link, while ${{R''_{{\rm{SD}}}}}$ only denotes the throughput of the S$-$D link. Accordingly, here we only need to consider ${{R''_{{\rm{SD}}}}} > {{R''_{{\rm{SR}}}}}$. In this case, we have ${h_{{\rm{SD}}}} > {h_{{\rm{SR}}}}$ and $t = {R''_{{\rm{SR}}}}$.

Based on the above discussion,  ${\mathcal{P  }_5}$ can be converted into
\begin{align}\notag
{\mathcal{P}_{5.2}}\;\;&\mathop {\max }\limits_{{P_{\rm{0}}},{P_1},\rho} {\rm{   }}{{R''_{{\rm{SD}}}}}\\\label{41}
{\rm{s.t.}}\;\;\;\;&{\rm{C1}}\! -\! 1,{\rm{ C6}},{\rm{ C7}}\!-\! 1,{\rm{ C10}}.
\end{align}
For ${\mathcal{P}_{5.2}}$, it is easy to prove $P_1^ *  = 0$ maximizing the objective function. Combining it with C1$-$1, we have ${P_0} = \frac{P}{\rho }$, and substitute it into the objective function, resulting in
\begin{align}\notag
{\mathcal{P}_{5.2.1}}\;\;&\mathop {\max }\limits_{\rho} {{R'''_{{\rm{SD}}}}}\\
{\rm{s.t.}}\;\;\;\;&{\rm{ C10\!-\!3}}:\frac{P}{{{P_{\max }}}} \le \rho  \le \min \left( {1,\frac{{P\eta {\left| {{h_{{\rm{SR}}}}} \right|^2}}}{{{P_{\rm{c}}}}}} \right),
\end{align}
where ${{{R'''_{{\rm{SD}}}}}}\!{\rm{ = }}{T_s}\!W\!\!\rho{\log _2}\!\left(\!{B\!\!+\!\!\frac{{P{\left|  {{h_{{\rm{SR}}}}} \right|^2}\!{\left| {{h_{{\rm{SD}}}}} \right|^2}}}{{\rho W\!{\sigma ^2}}}} \!\right)$. In terms of C10$-\!$3 in
 ${\mathcal{P}_{5.2.1}}$, the feasible region of $\rho$ can be determined as follows. Firstly, we drive $\frac{{{P_{\rm{c}}}}}{{\eta {\left| {{h_{{\rm{SR}}}}} \right|^2}}}\!\!\le \!\!{P_0}\! \! \le\!\! {P_{\max }}$ through combining C6 with C7$\! -\! $1 in ${\mathcal{P}_{5.2}}$. Next, we obtain C10$-\!$3 by substituting ${P_0}\!\!=\!\!\frac{P}{\rho }$ into the above inequality and combining C10.

Clearly, the constraint is linear and thus ${\mathcal{P}_{5.2.1}}$ can be solved by analyzing the objective function. The objective function can be written as  $q\!\left( \rho  \right){\rm{ \!= }}{T_s}\!W\!\rho {\log _2}\!\left( \!{B \! \!+ \!\! \frac{{P{\left| {{h_{{\rm{SR}}}}} \right|^2}{\left| {{h_{{\rm{SD}}}}} \right|^2}}}{{\rho W{\sigma ^2}}}}\right)$. It can
be observed that the second derivative of $q\left( \rho\right)$, i.e., $q''\left( \rho  \right){\rm{ = }}$
$\frac{{ - {T_s}\!W\!{P^2}{\left| {{h_{{\rm{SR}}}}} \right|^4}{\left| {{h_{{\rm{SD}}}}} \right|^4}}}{{\rho {{\left( {B\rho W{\sigma ^2} + P{\left| {{h_{{\rm{SR}}}}} \right|^2}{\left| {{h_{{\rm{SD}}}}} \right|^2}} \right)}^2}\ln 2 }}$, is negative, and thus the first
derivative of $q\left( \rho \right)$, i.e., $q'\left( \rho  \right){\rm{= }}{T_s}W{\log _2}\! \left( \! {B\!  +\!  \frac{{P{\left| {{h_{{\rm{SR}}}}} \right|^2}{\left| {{h_{{\rm{SD}}}}} \right|^2}}}{{\rho W {\sigma ^2}}}} \! \right) \! - \! $
$ \frac{{{T_s}WP{\left| {{h_{{\rm{SR}}}}}
\right|^2}{\left| {{h_{{\rm{SD}}}}} \right|^2}}}{{\left( {B\rho W {\sigma ^2} + P{\left| {{h_{{\rm{SR}}}}} \right|^2}{\left| {{h_{{\rm{SD}}}}} \right|^2}} \right)\ln 2}}$ is a monotonically decreasing function of $\rho$. Hence, the objective function is concave and there exists only one unique optimal solution in ${\mathcal{P}_{5.2.1}}$. Based on this, we can find the optimal time allocation factor by using the monotonicity or bisection method, as summarized in Algorithm 2, shown at the top of this page.
\begin{algorithm}[!t]
\setstretch{1}
    \caption{Algorithm for Solving ${\mathcal{P}_{5.2}}$}
    \label{alg:1}
    \renewcommand{\algorithmicrequire}{\textbf{Input:}}
    \renewcommand{\algorithmicensure}{\textbf{Output:}}
    \begin{algorithmic}[1]
    	\REQUIRE Coordinates of the IoT node, the HAP and the receiver
		\ENSURE ${{\rho ^*}}$, ${{\beta ^ *}}$, ${P_0^ *}$
        \IF{$q'\left( {\min \left( {1,\frac{{P\eta {\left| {{h_{{\rm{SR}}}}} \right|^2}}}{{{P_{\rm{c}}}}}} \right)} \right) \ge 0$}
\STATE ${\rho ^*} = \min \left( {1,\frac{{P\eta {\left| {{h_{{\rm{SR}}}}} \right|^2}}}{{{P_{\rm{c}}}}}} \right)$, $P_0^* = \frac{P}{{{\rho ^*}}}$;
\ELSIF{$q'\left( {\frac{P}{{{P_{\max }}}}} \right) \le 0$}
\STATE ${\rho ^*} = \frac{P}{{{P_{\max }}}}$, $P_0^* = \frac{P}{{{\rho ^*}}}$;
\ELSE
\STATE Obtain ${\rho ^*}$ via bisection method, $P_0^* = \frac{P}{{{\rho ^*}}}$;
\ENDIF
    \end{algorithmic}
\end{algorithm}
%

Until now, we have solved ${\mathcal{P}_{5.1}}$ and ${\mathcal{P}_{5.2}}$. From the above discussion,  we know that if ${h_{{\rm{SD}}}} \!\!\le \!\! {h_{{\rm{SR}}}}$, ${\mathcal{P}_{5}}$ is reduced to ${\mathcal{P}_{5.1}}$; Otherwise, ${\mathcal{P}_{5}}$ is equivalent to ${\mathcal{P}_{5.2}}$. Combining the above methods to solve  ${\mathcal{P}_{5.1}}$ and ${\mathcal{P}_{5.2}}$, we can easily solve ${\mathcal{P}_{5}}$.

Through the above analysis, we can obtain the optimal continuous solution denoted as $\left\{ {P_0^ * ,P_1^ * ,{\rho ^ * }} \right\}$, through which we can compute the optimal number of the subframes allocated to both phases, i.e., ${M^ * } = {\rho ^ * }L$, ${N^ * } = L - {M^ * }$. However, $\left\{ {{M^ * },{N^ * }} \right\}$ is a continuous solution, which cannot be used in practice. Therefore, we need to implement integer conversion so that $\left\{ {{M^ * },{N^ * }} \right\}$ can be converted to a feasible integer solution, i.e., $\left\{ {{M^\dag },{N^\dag }} \right\}$, satisfying the original problem, and then re-optimize $\left\{ {P_0 ,P_1 } \right\}$ by substituting $\left\{ {{M^\dag },{N^\dag }} \right\}$ into the original problem. However, the conventional integer conversion method, e.g., rounding, may not guarantee that the results meet the constraints of the original problem, since such a method has not considered the constraints of our formulated problem. Hence, in the \emph{third step}, we propose an integer conversion strategy, through which the derived integer solution meets all constraints of the original optimization problem. In what follows, we focus on designing an integer conversion strategy. Once $\left\{ {{M^\dag },{N^\dag }} \right\}$ is determined, $\left\{ {{P_0},{P_1}} \right\}$ can be derived by re-optimizing or re-computing.

\begin{algorithm}[!t]
\setstretch{1}
    \caption{Throughput Maximization based Algorithm}
    \label{alg:1}
    \renewcommand{\algorithmicrequire}{\textbf{Input:}}
    \renewcommand{\algorithmicensure}{\textbf{Output:}}
    \begin{algorithmic}[1]
    	\REQUIRE Coordinates of the IoT node, the HAP and the receiver
		\ENSURE ${{M^\dag}}$, ${{N^\dag}}$, ${{\beta ^\dag}}$, ${P_0^ \dag}$, ${P_1^\dag }$
        \IF{${h_{{\rm{SD}}}} \le {h_{{\rm{SR}}}}$}
\STATE Obtain $\left\{ {P_0^*,P_1^*,{\rho ^*},{\beta ^*},{t ^*}} \right\}$ by solving ${\mathcal{P}_{5.1}}$ by Algorithm 1;
        \ELSE
\STATE Obtain $\left\{ {P_0^*,P_1^*,{\rho ^*},{\beta ^*},{t^*}} \right\}$ by solving ${\mathcal{P}_{5.2}}$ by Algorithm 2;
\ENDIF
        \STATE Obtain discrete optimal solution $\left\{ {{M^\dag },{N^\dag }} \right\}$ by preforming integer conversion strategy;
        \IF{${h_{{\rm{SD}}}} \le {h_{{\rm{SR}}}}$}
        \STATE Obtain $\left\{ {{P_0^\dag },{P_1^\dag }} \right\}$ by re-optimizing $\left\{ {P_0,P_1 } \right\}$ and compute ${\beta ^\dag } = 1 - \frac{{{P_{\rm{c}}}}}{{\eta P_0^\dag {\left| {{h_{{\rm{SR}}}}} \right|^2}}}$;
 \ELSE
\STATE Re-compute $\left\{ {P_0^\dag,P_1^\dag ,{\beta ^\dag }} \right\}$ by $P_0^\dag  = \frac{{PL}}{{{M^ \dag }}}$, $P_1^ \dag  = 0$, ${\beta ^ \dag } = 1 - \frac{{{P_{\rm{c}}}}}{{\eta P_0^ \dag {\left| {{h_{{\rm{SR}}}}} \right|^2}}}$;
\ENDIF

    \end{algorithmic}
\end{algorithm}

In this work, the time-sharing technology is applied to convert a discrete optimization problem, i.e., ${\mathcal{P  }_3}$, to a continuous one, i.e., ${\mathcal{P  }_4}$, and our goal is to design an integer conversion strategy to obtain a feasible integral solution satisfying all the constraints of ${\mathcal{P  }_3}$. Due to the complexity of constraints in ${\mathcal{P  }_3}$ and the equivalence of ${\mathcal{P  }_3}$ and ${\mathcal{P  }_1}$ under the optimal solutions, we design an integer conversion strategy, through which the derived integral solution should satisfy all the related constraints with $M$ and $N$ of ${\mathcal{P  }_1}$, i.e., C1, C3, and C4.
Both C3 and C4 can be easily satisfied because we obtain ${{N^\dag }}$ via $L - {M^\dag }$ and $\rho  \in \left[ {0,1} \right]$. Thereby, the integer conversion strategy only depends on C1. We can find that the left side of C1 increases with $M$ when ${P_{\rm{0}}}> {P_1}$ and decreases with $M$ when ${P_{\rm{0}}}< {P_1}$. Combining it with the fact that the equality of C1 must be satisfied for maximizing the throughput of
the IoT node,
we have $M^\dag=\left\lfloor {{M^*}} \right\rfloor$ when $P_{\rm{0}}^* > {P_1}^*$ and
$M^\dag=\left\lceil {{M^*}} \right\rceil$ when $P_{\rm{0}}^* <{P_1}^*$, where $\left\lfloor  \cdot  \right\rfloor$ and $\left\lceil  \cdot  \right\rceil$ are the integer floor operation and integer ceil operation respectively.
Besides, the monotonicity of the left side in C1 is not related to $M$ when $P_0= P_1$. In this case, how to design $M^\dag $ only depends on the relationship of the objective function under $M^\dag=\left\lfloor {{M^*}} \right\rfloor$ and $M^\dag=\left\lceil {{M^*}} \right\rceil$.
Specifically, there are two cases for the relationship of the objective function, i.e., $\left\{ {P_{\rm{0}}^* = {P_1}^*} \right\}{\rm{\& }}\left\{ {{{\rm{R}}_{{\rm{sum}}}}\left( {\left\lfloor {{{\rm{M}}^{\rm{*}}}} \right\rfloor } \right) \ge {{\rm{R}}_{{\rm{sum}}}}\left( {\left\lceil {{{\rm{M}}^{\rm{*}}}} \right\rceil } \right)} \right\}$ (also denoted as Condition 1) and $\left\{ {P_{\rm{0}}^* = {P_1}^*} \right\}{\rm{\& }}$
$\left\{ {{{\rm{R}}_{{\rm{sum}}}}\left( {\left\lfloor {{{\rm{M}}^{\rm{*}}}} \right\rfloor } \right) < {{\rm{R}}_{{\rm{sum}}}}\left( {\left\lceil {{{\rm{M}}^{\rm{*}}}} \right\rceil } \right)} \right\}$ (also denoted as Condition 2).
Accordingly, the integer conversion strategy can be described as
\begin{align}\label{41}
{M^\dag }= \left\{ {\begin{array}{*{20}{c}}
{\left\lfloor {{M^*}} \right\rfloor ,P_{\rm{0}}^* > {P_1}^*{\rm{\;\;or\;\;Condition\;\;1}},}\\
{\left\lceil {{M^*}} \right\rceil ,P_{\rm{0}}^* < {P_1}^*{\rm{\;\;or\;\;Condition\;\;2}}.}
\end{array}} \right.
\end{align}
 Using \eqref{41}, the optimal integral solution of ${{N^\dag }}$ can be calculated by ${N^\dag } = L - {M^\dag }$.

After obtaining ${{M^\dag }}$ and ${{N^\dag }}$, we need to further explore the optimal transmit power of the HAP, i.e., $P_0$ and $P_1$, by substituting ${{M^\dag }}$ and ${{N^\dag }}$ into ${\mathcal{P}_{5.1}}$ or ${\mathcal{P}_{5.2}}$.

For \emph{Case 1 with ${h_{{\rm{SD}}}} \le {h_{{\rm{SR}}}}$}, we obtain the optimal transmit power of the HAP by solving ${\mathcal{P}_{6}}$,
given by
\begin{align}\notag
{\mathcal{P}_6}\;\;&\mathop {\max }\limits_{{P_0},{P_1},t} {\rm{   }}t\\\notag
{\rm{s.t.}}\;\;\;\;&{\rm{C1}}\! -\!5:\frac{{{M^\dag }{P_0}}}{L} + \frac{{{N^\dag }{P_1}}}{L} \le P,\\\notag
&{\rm{C6}},{\rm{C7}}\!- \!1 ,  \\\notag
&{\rm{C8}} \!- \!6:{{\hat R}_{{\rm{SR}}}} \ge t,\\\label{42}
&{\rm{C9}}\! - \!7:{{\hat R}_{\rm{D}}} \ge t,
\end{align}
where
${{\hat R}_{{\rm{SR}}}} \!\!=\!\! \frac{{{M^ \dag\! }{T_s}\!W}}{L}\!{\log _2}\!\left(\! \!{A \!+ \! \! \frac{{{P_0}{\left| {{h_{{\rm{SR}}}}} \right|^4}}}{{W{\sigma ^2}}}} \!\right)$, ${{\hat R}_{\rm{D}}}$ is shown as the top of the next page.
\begin{figure*}
\begin{align}
{{\hat R}_{\rm{D}}} = \left\{ {\begin{array}{*{20}{c}}
{\frac{{{N^\dag }{T_s}W}}{L}{{\log }_2}\left( {B + \frac{{{P_0}{{\left| {{h_{{\rm{SR}}}}} \right|}^2}{{\left| {{h_{{\rm{SD}}}}} \right|}^2} + {P_1}{{\left| {{h_{{\rm{RD}}}}} \right|}^2}}}{{W{\sigma ^2}}}} \right) + {T_s}W\left( {\frac{{2{M^\dag }}}{L} - 1} \right){{\log }_2}\left( {B + \frac{{{P_0}{{\left| {{h_{{\rm{SR}}}}} \right|}^2}{{\left| {{h_{{\rm{SD}}}}} \right|}^2}}}{{W{\sigma ^2}}}} \right),{\rm{   }}{M^\dag } \ge {N^\dag }}\\
\!\!\!\!\!\!\!\!\!\!\!\!\!\!\!\!\!\!\!\!\!\!\!\!
\!\!\!\!\!\!\!\!\!\!\!\!\!\!\!\!\!\!\!\!\!\!\!\!
\!\!\!\!\!\!\!\!\!\!\!\!\!\!\!\!\!\!\!\!\!\!\!\!
\!\!\!\!\!\!\!\!\!\!\!\!\!\!\!\!\!\!\!\!\!\!\!\!\!\!\!
{\frac{{{M^\dag }{T_s}W}}{L}{{\log }_2}\left( {B + \frac{{{P_0}{{\left| {{h_{{\rm{SR}}}}} \right|}^2}{{\left| {{h_{{\rm{SD}}}}} \right|}^2}}}{{W{\sigma ^2}}} + \frac{{{N^\dag }{P_1}{{\left| {{h_{{\rm{RD}}}}} \right|}^2}}}{{{M^\dag }W{\sigma ^2}}}} \right),{\rm{   }}{M^\dag } < {N^\dag }{\rm{                                                           }}}
\end{array}} \right.
\end{align}
\hrulefill
\end{figure*}

Due to the linear objective function and convex constraints, ${\mathcal{P}_6}$ is convex and can be directly solved by CVX.

For \emph{Case 2 with ${h_{{\rm{SD}}}} > {h_{{\rm{SR}}}}$}, the optimal transmit
power of the HAP can be re-calculated as $P_0^\dag  = \frac{{PL}}{{{M^\dag }}}$, $P_1^ \dag = 0$.

Combining the above three conversion steps, we proposed a throughput maximization-based algorithm, as summarized in  Algorithm 3.
Please note that after modifications, our proposed algorithm can also be adapted in the multi-IoT nodes scenario. Here we omit the detailed procedure due to the limited space.

\subsection{Complexity Analysis}
In this subsection, we evaluate the computational complexity of Algorithm 3. Firstly, we analyze the computational complexities of Algorithm 1 and Algorithm 2. For Algorithm 1, the interior point method \cite{8675436,9159909} is adopted and its computational complexity is $\mathcal{O}\left( {\sqrt {{K}} \log \left( {{K}} \right)} \right)$ for each iteration, where ${{K}}$ denotes the number of the inequality constraints of optimization problem. Assume that the maximum number of iterations of outer layer for solving ${\mathcal{P  }_{5.1.3}}$ is $K_1$, then the computation complexity for Algorithm 1 can be calculated as ${K_1}\mathcal{O}\left( {\sqrt {{K_2}} \log \left( {{K_2}} \right)} \right)+\mathcal{O}\left( {\sqrt {{K_3}} \log \left( {{K_3}} \right)} \right)$, where ${K_2}$ and ${K_3}$ denote the number of the inequality constraints of ${\mathcal{P  }_{5.1.3}}$ and ${\mathcal{P  }_{5.1.5}}$, i.e., ${K_2}\!\!=\!\!{K_3}\!\!=\!\!7$, respectively. For Algorithm 2, the worse case is to adopt the bisection method. For the bisection method \cite{9179779}, the searching range is set as $\left[ {\frac{P}{{{P_{\max }}}},\min \left( {1,\frac{{P\eta {\left| {{h_{{\rm{SR}}}}} \right|^2}}}{{{P_{\rm{c}}}}}} \right)} \right]$, which will reduce by half after each iteration, and thus the number of iterations is $K_4= {\log \left( {\frac{{{P \mathord{\left/
 {\vphantom {P {{P_{\max }}}}} \right.
 \kern-\nulldelimiterspace} {{P_{\max }}}} - \min \left( {1,{{P\eta {\left| {{h_{{\rm{SR}}}}} \right|^2}} \mathord{\left/
 {\vphantom {{P\eta {\left| {{h_{{\rm{SR}}}}} \right|^2}} {{P_{\rm{c}}}}}} \right.
 \kern-\nulldelimiterspace} {{P_{\rm{c}}}}}} \right)}}{\varepsilon }} \right)}$, where $\varepsilon $ denotes the accuracy. Therefore, the computational complexity of the bisection method can be expressed as ${\mathcal{O}\left( K_4 \right)}$, which equals the complexity of Algorithm 2. Subsequently, we analyze the computational complexity of re-optimizing or re-computing $\left\{ {P_0,P_1 } \right\}$ after obtaining the discrete subframe allocation. The computational complexity of the worst scenario, i.e., re-optimizing $\left\{ {P_0,P_1 } \right\}$ when ${h_{{\rm{SD}}}} \le {h_{{\rm{SR}}}}$, is $\mathcal{O}\left( {\sqrt {{K_5}} \log \left( {{K_5}} \right)} \right)$ by using the interior point method, where $K_5$ represents the number of the inequality constraints of ${\mathcal{P  }_6}$, i.e., $K_5=6$. Based on the above analysis, the total computational complexity of Algorithm 3 can be calculated as $\max \left\{ {{K_1}{\cal O}\left( {\sqrt {{K_2}} \log \left( {{K_2}} \right)} \right) + {\cal O}\left( {\sqrt {{K_3}} \log \left( {{K_3}} \right)} \right),{\cal O}\left( {{K_4}} \right)} \right\} + {\cal O}\left( {\sqrt {{K_5}} \log \left( {{K_5}} \right)} \right)$.
\section{Simulation}
In this section, we present numerical simulation results to verify the effectiveness and the superiority of our proposed resource allocation scheme for the relay-enabled BC network.
In this network, the IoT node, the HAP and the receiver are located within a 2-D region of 100$ \times$100 m${^2}$ and described with a 2-D Cartesian coordinate system.
The basic simulation parameters considered in this paper are shown in Table I \cite{9449060}. Besides, the channel gains are model as ${\left| {{h_{{\rm{SD}}}}} \right|^2} = {\xi _{{\rm{SD}}}}d_{{\rm{SD}}}^{ - {\alpha _1}}$, ${\left| {{h_{{\rm{SR}}}}} \right|^2} = {\xi _{{\rm{SR}}}}d_{{\rm{SR}}}^{ - {\alpha _2}}$ and ${\left| {{h_{{\rm{RD}}}}} \right|^2}= {\xi _{{\rm{RD}}}}d_{{\rm{RD}}}^{ - {\alpha _3}}$, where ${\xi _i}$ and ${d_i}$ denote the power gain of the small-scale fading and the distance for the $i$th link $i \in \left\{ {{\rm{SD}},{\rm{SR}},{\rm{RD}}} \right\}$, and ${\alpha _1}$, ${\alpha _2}$ and ${\alpha _3}$ represent the path loss exponents of the S$-$D, S$-$R and R$-$D links, respectively. Unless otherwise specified, we set ${\alpha _2}=2.7$ and ${\alpha _3}=2.7$. In what follows, we analyze the simulation results from the following aspects: 1) Convergence of Algorithm 3; 2) Performance analysis of the time-sharing technology; 3) Comparison with other schemes; 4) Impacts of network parameters on throughput and subframe allocation.
\begin{table}[!t]
\centering
\caption*{{Table I. Simulation Parameters}}
\begin{tabular}{|l|l|}
\hline

Simulation parameter& Value\\
\hline
Coordinate of the IoT node S&$\left( {0,0} \right)$\\
\hline
Coordinate of the HAP R& $\left( {20} \right.$m$,20$m$\left. {} \right)$\\
\hline
Coordinate of the receiver D& $\left( {100} \right.$m$,0$$\left. {} \right)$\\
\hline
The duration of one frame $T_s$& 10 ms\\
\hline
Channel bandwidth $W$ & 10 kHz\\
\hline
Noise power spectral density ${\sigma ^2}$& $-$100 dBm/Hz\\
\hline
Energy conversion efficiency $\eta $&0.5\\
\hline
Backscatter circuit power consumption ${P_{\rm{c}}}$& 200 $\mu$W\\
\hline
Total energy in each frame of the HAP $E$&200 mW \\
\hline
\end{tabular}
\end{table}
\subsection{Convergence of Algorithm 3}
Fig. 2 shows the convergence of the proposed Algorithm 3 from the perspective of iteration variables, i.e., $\rho$ and $a$, versus the iteration number. We plot the convergence curves under different parameters, i.e., $\alpha_1\!\!=\!\!2.5$, $P_{\rm{max}}\!\!=\!\!20$W and $\alpha_1\!\!=\!\!2.5$, $P_{\rm{max}}\!\!=\!\!30$W. It can be observed that less than three iterations for $P_{\rm{max}}\!\!=\!\!30$W and just only one iterations for $P_{\rm{max}}\!\!=\!\!20$W are required for Algorithm 3 to converge a value. That illustrates that the proposed algorithm has a quick convergence. Besides, it can be clearly seen that the convergence speed of $P_{\rm{max}}\!\!=\!\!20$W is faster than that of $P_{\rm{max}}\!\!=\!\!30$W. This phenomenon can be interpreted that the convergence performance of SCA algorithm largely depends on the setting of the initial value. However, the initial value setting, i.e., $\rho^0\!\!=\!\!0.7$, $a^0\!\!=\!\!6$, may not be suitable for $P_{\rm{max}}\!\!=\!\!30$W compared with $P_{\rm{max}}\!\!=\!\!20$W, leading to more iterations.
\begin{figure}[t]
  \centering
  \includegraphics[width=0.34\textwidth]{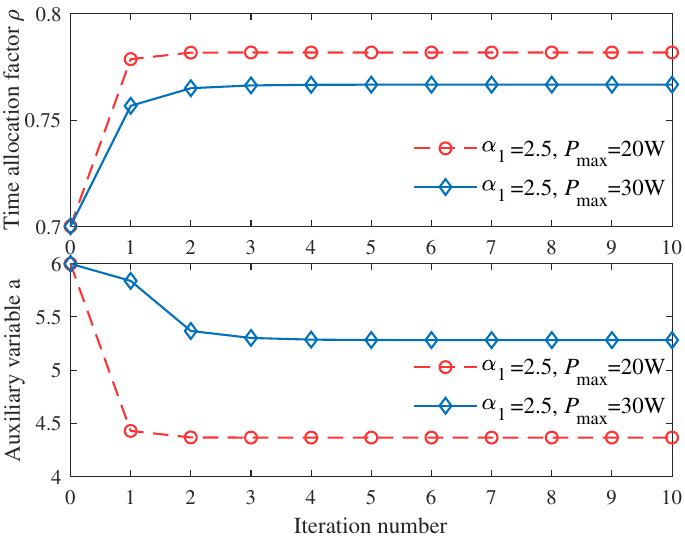}  \\
  \caption*{Fig. 2. \small{ Iteration variables versus iteration number.}}\label{fig2}
\end{figure}

\subsection{Performance Analysis of the
Time-Sharing Technology}
In this subsection, we would like to analyze the gap caused by using the time-sharing technology. Fig. 3 shows the comparisons between the proposed optimization algorithm and the exhaustive search algorithm in terms of achievable throughput under the setting of $P_{\rm{max}}\!\!=\!\!20$W, $L\!\!=\!\!20$ and $P_{\rm{max}}\!\!=\!\!30$W, $L\!\!=\!\!20$. Note that the exhaustive search algorithm can obtain global optimal solution for our formulated problem, at the cost of a high computational complexity. It can be observed that the gaps between the exhaustive search algorithm and our proposed Algorithm 3 is obvious when the number of subframes is few.
This phenomenon may be caused by the following two reasons. On the one hand, the integer conversion strategy needs to be performed after obtaining the optimal continuous solutions through the time-sharing technology, which may lead to the suboptimal discrete solutions in order to meet the constraints of the original optimization problem. On the other hand, for the time-sharing technology, the gap can be ignored when the number of subframes is sufficiently large. This indicates that when the number of subframes tends to infinity, the gap caused by the time-sharing technology is zero. This is because when the number of subframes tends to infinity, $N/L$ and $M/L$ can be arbitrary variables in the range 0 to 1. To illustrate this conclusion, Fig. 4 shows the gap caused by the time-sharing technology versus the number of total subframes. Let's pick out some points with large gaps in Fig. 3, i.e., $\alpha_1\!\!=\!\!4.0$ and $\alpha_1\!\!=\!\!3.9$ in $P_{\rm{max}}\!\!=\!\!20$W, as well as $\alpha_1\!\!=\!\!2.6$ and $\alpha_1\!\!=\!\!2.9$ in $P_{\rm{max}}\!\!=\!\!30$W.
\begin{figure}[t]
  \centering
  \includegraphics[width=0.35\textwidth]{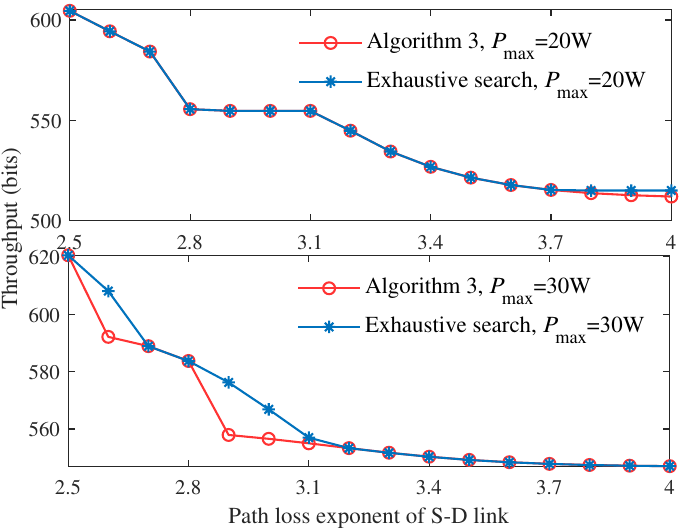}  \\
  \caption*{Fig. 3. \small{Throughput under different algorithms and maximum transmit power versus the path loss exponent of S$-$D link.}}\label{fig3}
\end{figure}
\begin{figure}[t]
  \centering
  \includegraphics[width=0.35\textwidth]{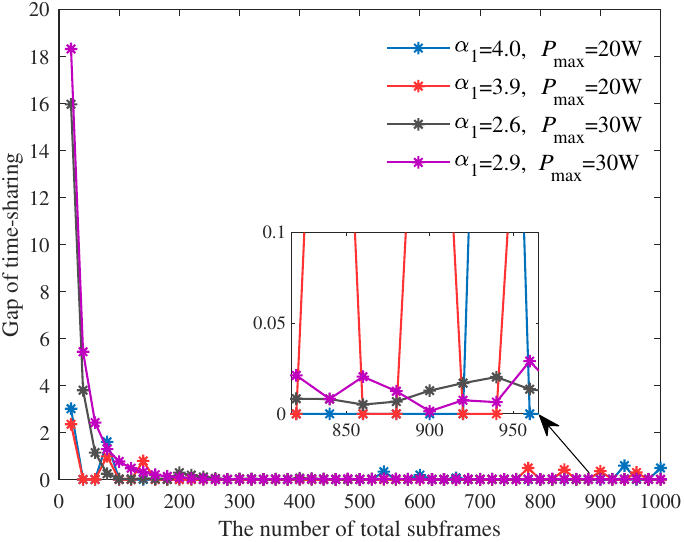}  \\
  \caption*{Fig. 4. \small{Gap caused by the time-sharing technology versus the number of total subframes.}}\label{fig4}
\end{figure}
 \!\!We can see that the gap between the above two algorithms tends to zero with the increase of the number of total subframes on the whole, although there may be some fluctuations.
 The reason for fluctuations is as follows. After adopting the time-sharing technology, the integer conversion is applied to convert the obtained optimal continuous time into a discrete one.
 When $\alpha_1\!\!=\!\!4.0$ and $\alpha_1\!\!=\!\!3.9$ in $P_{\rm{max}}\!\!=\!\!20$W, the optimal power allocation for ${{\cal P}_{5.1}}$ satisfies ${P_{\rm{0}}^* = {P_1}^*}$, thus the integer conversion strategy is determined by Condition 1 and Condition 2.
 When $\alpha_1\!\!=\!\!2.6$ and $\alpha_1\!\!=\!\!2.9$ in $P_{\rm{max}}\!\!=\!\!30$W, the optimal power allocation for ${{\cal P}_{5.1}}$ satisfies ${P_{\rm{0}}^* >{P_1}^*}$, the integer floor operation needs to be performed to obtain the optimal integral solution.
 Therefore, in order to meet the constraints
of the original optimization problem, the corresponding integer conversion strategy for our formulated problem needs to be followed, which may lead to the suboptimal discrete solutions
resulting in fluctuations at some points.
 We can also find that the fluctuations under
      $\alpha_1\!\!=\!\!4.0$ and $\alpha_1\!\!=\!\!3.9$ in $P_{\rm{max}}\!\!=\!\!20$W seem to be more extensive than that under $\alpha_1\!\!=\!\!2.6$ and $\alpha_1\!\!=\!\!2.9$ in $P_{\rm{max}}\!\!=\!\!30$W  when the total number of subframes is large, so that there seems no fluctuation in the curve under $\alpha_1\!\!=\!\!2.6$ and $\alpha_1\!\!=\!\!2.9$ in $P_{\rm{max}}\!\!=\!\!30$W, such phenomenon can be interpret as follows.
When $\alpha_1\!\!=\!\!4.0$ and $\alpha_1\!\!=\!\!3.9$ in $P_{\rm{max}}\!\!=\!\!20$W, the optimal power allocation for ${{\cal P}_{5.1}}$ and ${{\cal P}_{6}}$ all satisfies ${P_{\rm{0}}^* = {P_1}^*}=20W$, which means that the gap caused by the integer conversion strategy has not been compensated through the further optimization.
However, the optimal power allocation for ${{\cal P}_{5.1}}$ and ${{\cal P}_{6}}$ are not equal when $\alpha_1\!\!=\!\!2.6$ and $\alpha_1\!\!=\!\!2.9$ in $P_{\rm{max}}\!\!=\!\!30$W, thus the gap caused by the integer conversion strategy has been compensated partly through the further optimization.

\subsection{Comparison with Other Schemes}
For comparison between the proposed scheme and other schemes, we will do it from the following two aspects. Firstly, we compare the proposed scheme with the following three fixed time
schemes: 1) BC; 2) relay-enabled BC; 3) opportunistic relay-enabled BC. Besides, we compare the proposed scheme with other related schemes, which optimize the continuous and unequal time variables. Fig. 5 plots the throughput versus the path loss exponent of S$-$D link of different schemes under $\alpha_2=3.2$ and ${\rm{D}} = \left\{ {50,0} \right\}$. We can see that the throughput achieved by all the schemes will decrease with the increase of the path loss exponent of S$-$D link. It can also be observed that the proposed scheme always outperforms the other schemes in terms of achievable throughput. The reasons are summarized as follows. The fixed time scheme is considered in other schemes without the optimization of time allocation, while the proposed scheme in this paper optimizes the unequal time allocation, which takes full advantage of changes in network parameters. Fig. 6 shows the comparison of our proposed scheme and other related schemes considering the optimization of the continuous time variables. In this figure, we plot two curves of our proposed scheme under different number of total subframes, i.e., $L=20$ in reality and $L=1000$ in simulation, respectively. We can clearly see that the achievable throughput of our proposed scheme under setting $L=20$ is less than obtained from other related scheme. There exists two reasons as follows. On the one hand, other related schemes apply an upper bound of the accurate system-throughput expression we obtained in this paper to find the optimal resource allocation scheme. On the other hand, other related schemes can obtain more ideal throughput by optimizing the continuous time, while this is infeasible in reality.
Comparing the other curve under $L=1000$ with other related schemes, the achievable throughput of our proposed scheme is less than other related schemes in the whole because an upper bound of the accurate throughput expression is used in other related schemes, which confirms \emph{Remark 2}.
\begin{figure}[t]
  \centering
  \includegraphics[width=0.34\textwidth]{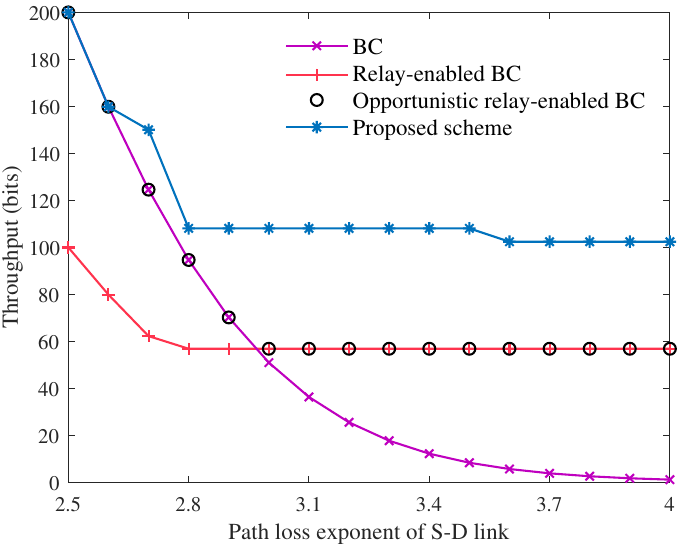}  \\
  \caption*{Fig. 5. \small{Throughput under different schemes versus the path loss exponent of S$-$D link.}}\label{fig5}
\end{figure}
\begin{figure}[t]
  \centering
  \includegraphics[width=0.34\textwidth]{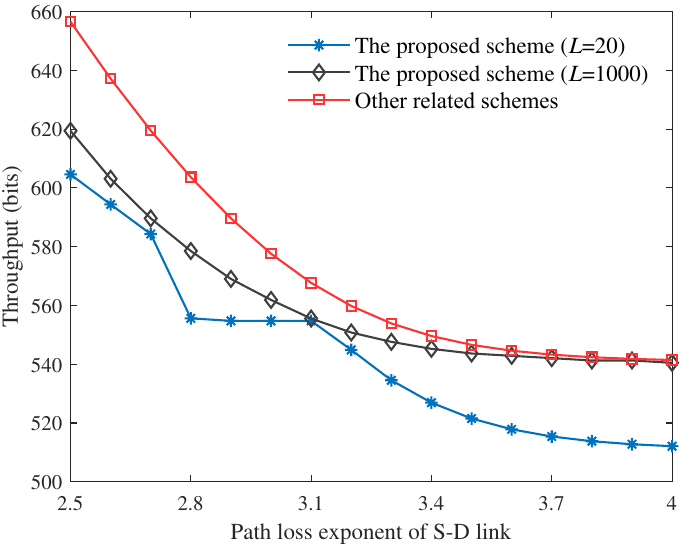}  \\
  \caption*{Fig. 6. \small{Throughput under the proposed scheme and other related schemes versus the path loss exponent of S$-$D link.}}\label{fig6}
\end{figure}

\subsection{Impacts of Network Parameters on Throughput and Time Allocation}
 Fig. 7 depicts the throughput versus the $\rm{X}-$coordinate and $\rm{Y}-$coordinate of the HAP under different setting of $P_{\rm{max}}$. We can see that the achievable throughput decreases with the increase of whatever $\rm{Y}-$coordinate or $\rm{X}-$coordinate of the HAP. The reason is as
follows. When $\rm{Y}-$coordinate of the HAP decreases, the distances both S$-$R and R$-$D links is shortened, a better throughput can be obtained. Besides, with the decrease of $\rm{X}-$coordinate of the HAP, the distance of the S$-$R link is shorten while the R$-$D link is increased, that leads to the uncertainty of throughput changes. However, the reduction of the achievable throughput caused by the increase of distance of R$-$D link can be compensated by optimizing the other network parameters. At the same time, the decrease of X$-$coordinate of the HAP not only shortens the distance of S$-$D link but also weakens the impact of the double path loss. Therefore, we can obtain a better throughput with the decrease of X$-$coordinate of the HAP. Fig. 8 shows the number of subframes allocated for each phase, i.e., the backscatter phase and the relay phase, versus the path loss exponent of S$-$D link. When the channel quality of S$-$D link is worse, the number of subframes allocated to the relay phase is more.

\begin{figure}[t]
  \centering
  \includegraphics[width=0.34\textwidth]{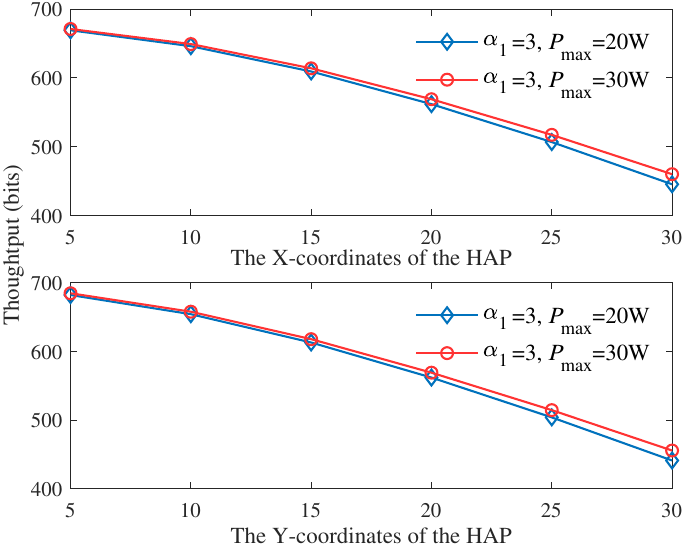}  \\
  \caption*{Fig. 7. \small{Throughput versus coordinate of the HAP.}}\label{fig7}
\end{figure}
\begin{figure}[t]
  \centering
  \includegraphics[width=0.34\textwidth]{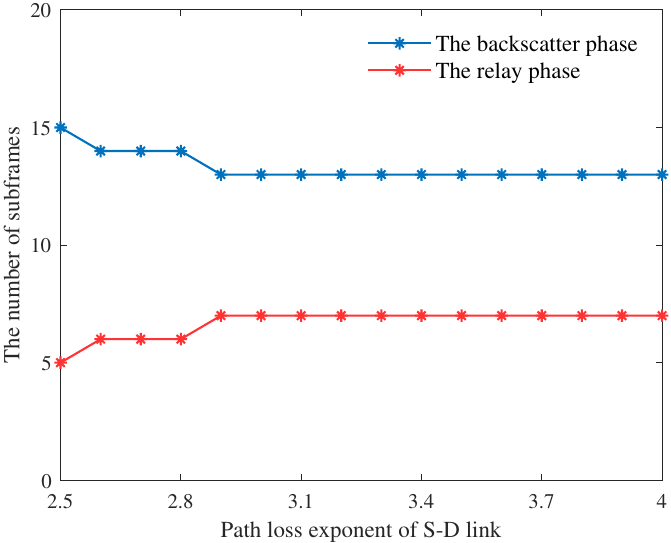}  \\
  \caption*{Fig. 8. \small{The number of subframes versus the path loss exponent of S$-$D link.}}\label{fig8}
\end{figure}

\section{Conclusion}
In this paper, considering the unequal and discrete time allocation, we proposed a practical resource allocation scheme for the relay-enabled BC network based on the linear mapping method.
Specifically, we derived the accuracy system-throughput expression by applying the linear mapping. Based on this, a mixed-integer non-convex optimization problem is formulated by jointly optimizing the linear mapping matrix, the power allocation of the HAP and the PRC of the IoT node to achieve the maximum throughput.
We proposed an iterative algorithm to solve this non-convex problem via introducing auxiliary variables, exploiting KKT conditions and time-sharing technology, and proposing an integral conversion strategy.
The complexity of the proposed algorithm was also analyzed. Finally, simulation results were provided to verify the convergence of the proposed algorithm, the superiority of the proposed scheme, and the impacts of network parameters on the system throughput.

\section*{Appendix A}
According to the auxiliary operation $t = \min \left\{ {{R_{{\rm{SR}}}},{{R_{\rm{D}}}}} \right\}$, we prove Lemma 1 from the following two cases, i.e., Case 1: ${R_{{\rm{SR}}}} \!\!\ge\! \!{R_{\rm{D}}}$, and Case 2: ${R_{{\rm{SR}}}}\!\! <\!\! {R_{\rm{D}}}$.
For Case 1, supposed that ${R_{\rm{D}}} \!\!> \!\!t^+$ in C9 and other constraints are satisfied for the optimal solutions $\left\{ {\lambda _1^ + , \cdots ,\lambda _{\min \left( {{M^ + },{N^ + }} \right)}^ + ,P_0^ + ,P_1^ + ,{\beta ^ + },{M^ + },} \right.$ $\left. {{N^ + },{t^ + }} \right\}$. However, there exists another solutions $\left\{ {\lambda _1^ + , \cdots ,} \right.$
$\left. {\lambda _{\min \left( {{M^ + },{N^ + }} \right)}^ + ,P_0^ + ,P_1^ + ,{\beta ^ + },{M^ + },{N^ + },{t^*}} \right\}$
 satisfying ${R_{\rm{D}}} \!\!=\!\! {t^ * }$ and other constraints when ${t^* } \ge {t^ + }$. At this time, the objective value keeps unchanged or increases since the objective function does not decrease with variable $t$. This means that the equality in C9 must hold for the optimal solutions under condition ${R_{{\rm{SR}}}} \ge {R_{\rm{D}}}$. It can be reduced to a special case of ${\mathcal{P}_1}$. By the same argument, we gain that the equality in C8 holds for the optimal solutions under condition ${R_{{\rm{SR}}}} < {R_{\rm{D}}}$. That denotes the other special case of ${\mathcal{P}_1}$. Therefore, The optimal solutions from ${\mathcal{P}_2}$ must satisfy one of constraint C8 or C9 with equality and equals that from ${\mathcal{P}_1}$.
\section*{Appendix B}
Let ${{\mu _1}}$, ${{\mu _2}}$, ${{\mu _3}}$, ${{\mu _4}}$, ${{\mu _5}}$ and ${{\mu _6}}$ denote the nonnegative Lagrange multipliers associated with C1, C2, C3, C5, C8, C9. Then the partial Lagrangian function of ${\mathcal{P}_2}$ is given by
\begin{align}\notag
&{\mathbb{L}} \!= \!\max \left\{ \! {\frac{{{T_s}\!W\!M}}{L}{{\log }_2}\!\left(\! {1 \!+\! \frac{{\theta _1}}{{W\!{\sigma ^2}}}} \!\right)\!,t\!} \right\} \!+ \! {\mu _1}\!\left( \!\!{E \!- \! \frac{{M\!{P_{\rm{0}}}\!{T_s}}}{L}\! - \! \frac{{N\!{P_{\rm{1}}}\!{T_s}}}{L}} \!\right)\\\notag
 &+ {\mu _2}\left( {\eta \left( {1 - \beta } \right){P_0}{\left| {{h_{{\rm{SR}}}}} \right|^2} - {P_{\rm{c}}}} \right) + {\mu _3}\left( {M + N - L} \right)\\\notag
 &+ {\mu _4}\left( {N -\!\!\!\!\sum\limits_{i = 1}^{\min \left( {M,N} \right)}\!\!\!\!{{\lambda _i}} } \right) + {\mu _5}\left[ {\frac{{{T_s}\!W\!M}}{L}{{\log }_2}\left( {{\rm{1}} + \frac{{\theta _3}}{{W\!{\sigma ^2}}}} \right) - t} \right]\\\notag
 &+ {\mu _6}\left[ {\frac{{{T_s}\!W\!M}}{L}{{\log }_2}\left( {1 + \frac{{\theta _1}}{{W\!{\sigma ^2}}}} \right) - t} \right]\\\tag{B.1}
& + \frac{{{\mu _6}{T_s}\!W}}{L}\!\!\sum\limits_{i = 1}^{\min \left( {M,N} \right)}\!\!\!\!{{{\log }_2}} \left( {1 + \frac{{{\theta _2}/\left(W\!{\sigma ^2}\right)}}{{1 + {\theta _1}/\left(W\!{\sigma ^2}\right)}}{\lambda _i}} \right),
\end{align}
where ${\theta _1}\buildrel \Delta \over = \beta {P_0}{\left| {{h_{{\rm{SR}}}}} \right|^2}{\left| {{h_{{\rm{SD}}}}} \right|^2}$, ${\theta _2} \buildrel \Delta \over = {P_1}{\left| {{h_{{\rm{RD}}}}} \right|^2}$ and ${\theta _3}\buildrel \Delta \over = \beta {P_0}{\left| {{h_{{\rm{SR}}}}} \right|^4}$ for convenience. By taking the partial derivative of $\mathbb{L}$ with respect to ${{\lambda _i}}$, we have
\begin{align}\tag{B.2}
\frac{{\partial {\mathbb{L}}}}{{\partial {\lambda _i}}} = \frac{{{\mu _6}{T_s}\!W}}{{L}{\ln 2}}\left( {\frac{1}{{\left( {{\lambda _i} + \frac{{1 + {\theta _1}/\left(W\!{\sigma ^2}\right)}}{{{\theta _2}/\left(W\!{\sigma ^2}\right)}}} \right)}}} \right) - {\mu _4}.
\end{align}
By letting ${{\partial \mathbb{L}} \mathord{\left/{\vphantom {{\partial \mathcal{L}} {\partial {\lambda _i} = 0}}} \right.
\kern-\nulldelimiterspace} {\partial {\lambda _i} = 0}}$, we can derive the necessary condition of optimal eigenvalues.
\begin{align}\tag{B.3}
\lambda _i^ *  = \max \left\{ {0,\frac{{{T_s}\!W\!{\mu _6}}}{{L{\mu _4}\ln 2}} - \frac{{1 + {\theta _1}/\left(W\!{\sigma ^2}\right)}}{{{\theta _2}/\left(W\!{\sigma ^2}\right)}}} \right\}.
\end{align}
Since $\sum\limits_{i = 1}^{\min \left( {M,N} \right)} {{\lambda _i}}  = N$, the optimal eigenvalues are given by
\begin{align}\tag{B.4}
\lambda _i^ *  = \frac{{{T_s}\!W\!{\mu _6}}}{{L{\mu _4}\ln 2}} - \frac{{1 + {\theta _1}/\left(W\!{\sigma ^2}\right)}}{{{\theta _2}/\left(W\!{\sigma ^2}\right)}}.
\end{align}

It can be observed that there exist the unique solution satisfying KKT conditions for ${\mathcal{P}_2}$. Therefore, the optimal solution for problem ${\mathcal{P}_2}$ is ${\lambda _i} = \frac{N}{{\min \left( {M,N} \right)}}$ for $i = 1, \cdots ,{\min \left( {M,N} \right)}$. Lemma 2 is proven.

\begin{figure*}
\begin{align}\notag
&y'\left( \rho  \right) = - \frac{1}{{{\rho ^2}}}, {{f'_\rho }}\left( {\rho ,b} \right) = - \frac{b}{{{\rho ^2}}}, {{f'_b}}\left( {\rho ,b} \right) = \frac{1}{\rho }, {{f''_{\rho \rho }}}\left( {\rho ,b} \right)= \frac{{2b}}{{{\rho ^3}}},\\\notag
&{{g'_\rho} }\left( {\rho ,a} \right) = {T_s}W{\log _2}\left( {{\rm{A}} + \frac{{a{{\left| {{h_{{\rm{SR}}}}} \right|}^4}}}{{\left( {1 - \rho } \right)W{\sigma ^2}}}} \right) + \frac{{{T_s}W\rho }}{{\left( {\frac{{{\rm{A}}{{\left( {1 - \rho } \right)}^2}W{\sigma ^2}}}{{a{{\left| {{h_{{\rm{SR}}}}} \right|}^4}}} + \left( {1 - \rho } \right)} \right)\ln 2}}, {{g'_a}}\left( {\rho ,a} \right) = \frac{{{T_s}W\rho }}{{\left( {\frac{{{\rm{A}}\left( {1 - \rho } \right)W{\sigma ^2}}}{{{{\left| {{h_{{\rm{SR}}}}} \right|}^4}}} + a} \right)\ln 2}},\\\notag
&{{w'_\rho} }\left( {\rho ,a} \right) = 2{T_s}W{\log _2}\left( {B + \frac{{a{{\left| {{h_{{\rm{SR}}}}} \right|}^2}{{\left| {{h_{{\rm{SD}}}}} \right|}^2}}}{{\left( {1 - \rho } \right)W{\sigma ^2}}}} \right) + \frac{{{T_s}W\left( {2\rho  - 1} \right)}}{{\left( {\frac{{B{{\left( {1 - \rho } \right)}^2}W{\sigma ^2}}}{{a{{\left| {{h_{{\rm{SR}}}}} \right|}^2}{{\left| {{h_{{\rm{SD}}}}} \right|}^2}}} + \left( {1 - \rho } \right)} \right)\ln 2}},
{{w'_a}}\left( {\rho ,a} \right) = \frac{{{T_s}W\left( {2\rho  - 1} \right)}}{{\left( {\frac{{B\left( {1 - \rho } \right)W{\sigma ^2}}}{{{{\left| {{h_{{\rm{SR}}}}} \right|}^2}{{\left| {{h_{{\rm{SD}}}}} \right|}^2}}} + a} \right)\ln 2}},\\\tag{D.1}
&{{g''_{aa}}}\left( {\rho ,a} \right) =  - \frac{{{T_s}W\rho }}{{{{\left( {\frac{{{\rm{A}}\left( {1 - \rho } \right)W{\sigma ^2}}}{{{{\left| {{h_{{\rm{SR}}}}} \right|}^4}}} + a} \right)}^2}\ln 2}},
{{w''_{aa}}}\left( {\rho ,a} \right) =  - \frac{{{T_s}W\left( {2\rho  - 1} \right)}}{{{{\left( {\frac{{{\rm{B}}\left( {1 - \rho } \right)W{\sigma ^2}}}{{{{\left| {{h_{{\rm{SR}}}}} \right|}^2}{{\left| {{h_{{\rm{SD}}}}} \right|}^2}}} + a} \right)}^2}\ln 2}}.
\end{align}
\hrulefill
\end{figure*}

\section*{Appendix C}
First of all, we prove that the equality of constraint C1$-$1 must hold by means of contradiction. Specifically, we suppose that the optimal solution denoted $\left\{ {P_0^ + ,P_1^ + ,{\beta ^ + },{\rho^ + },{t^ + }} \right\}$ satisfies ${\rho ^ + }P_0^ +  + \left( {1 - {\rho ^ + }} \right)P_1^ +  < P$ and the remaining constraints. However, we can find another solution $\left\{ {P_0^ * ,P_1^ * ,{\beta ^ +},{\rho ^ + },{t^ * }} \right\}$ satisfying ${\rho ^ + }P_0^ *  + \left( {1 - {\rho ^ + }} \right)P_1^ * \! = \!P$ where $P_0^ *  > P_0^ + $, $P_1^ *  > P_1^ + $, and ${t ^ * } > {t^ + }$. At this time, other constraints are also satisfied and the objective function obtains a better value, which conflicts with the original assumption. Therefore, the equality of constraint C1$-$1 must hold for the optimal solutions.

Next, we prove that the equality of constraint C2$-$1 must hold still by means of contradiction. Specifically, assume that $\left\{ {P_0^ + ,P_1^ + ,{\beta ^ + },{\rho^ + },{t^ + }} \right\}$ is the optimal solution to ${\mathcal{P  }_4}$, where $\eta \!\left( {1 \!- \!{\beta ^ + }} \right)\!P_0^ + {\left| {{h_{{\rm{SR}}}}} \right|^2} \!\!> \!\!P_c$ always holds. However, we can construct another
solution $\left\{ {P_0^ + ,P_1^ + ,{\beta ^ * },{\rho^ + },{t^ *}} \right\}$ satisfying ${\beta ^ * } > {\beta ^ + }$, ${t ^ * } > {t^ + }$ and $\eta \left( {1 - {\beta ^ * }} \right)P_0^ * {\left| {{h_{{\rm{SR}}}}} \right|^2} \!=\! P_c$. Obviously, we can find that the constructed solution can achieve a better throughput and satisfy all the constraints of ${\mathcal{P  }_4}$, which conflicts with the original assumption. Based on this, the equality of constraint C2$-$1 must hold for the optimal solution and we can derive the optimal PRC, i.e.,  ${\beta ^ * } = 1 - \frac{{{P_{\rm{c}}}}}{{\eta {P_{\rm{0}}}{\left| {{h_{{\rm{SR}}}}} \right|^2}}}$.

The proof is completed.
\section*{Appendix D}
For convenience, we define the following functions to denote the complicated expressions in ${\mathcal{P}_{5.1.2}}$.
$y\left( \rho  \right) \buildrel \Delta \over = \frac{1}{\rho }$,
$f\left( {\rho ,b} \right)\buildrel \Delta \over =\frac{b}{\rho }$,
$g\left( {\rho ,a} \right)\buildrel \Delta \over =
{T_s}W\rho {\log _2}\left( {{\rm{A}} + \frac{{a{\left| {{h_{{\rm{SR}}}}} \right|^4}}}{{\left( {1 - \rho } \right)W{\sigma ^2}}}} \right)$,
$w\left( {\rho,a} \right)$
$\buildrel \Delta \over =
{T_s}\!W\!\left( {2\rho-1} \right){\log _2}\left( \!{B+\! \!\frac{{a{\left| {{h_{{\rm{SR}}}}} \right|^2}{\left| {{h_{{\rm{SD}}}}} \right|^2}}}{{\left( {1 - \rho } \right)W{\sigma ^2}}}} \!\right)$,
$e\left( {\rho ,a,b} \right) = {T_s}W$
$\left( {1 - \rho } \right){\log _2}\left( {B + \frac{{a{{\left| {{h_{{\rm{SR}}}}} \right|}^2}{{\left| {{h_{{\rm{SD}}}}} \right|}^2} + b{{\left| {{h_{{\rm{RD}}}}} \right|}^2}}}{{\left( {1 - \rho } \right)W{\sigma ^2}}}} \right)$.

In terms of these functions, we take the partial derivative with respect to $\rho$, $a$ and $b$, respectively, which are shown at the top of this page.

Thereby, we have
$y\left( \rho  \right) \ge {y_{lb}}\left( {{\rho ^j}} \right)$, $f\left( {\rho ,b} \right) \le {f_{ub}}\left( {{\rho ^j},{b^j}} \right)$, $g\left( {\rho ,a} \right) \ge {g_{lb}}\left( {{\rho ^j},{a^j}} \right)$ and $w\left( {\rho ,a} \right) \ge {w_{lb}}\left( {{\rho ^j},{a^j}} \right)$, where
\begin{align}\tag{D.2}
{y_{lb}}\left( {{\rho ^j}} \right) = y\left( {{\rho ^j}} \right) + y'\left( {{\rho ^j}} \right)\left( {\rho  - {\rho ^j}} \right){\rm{ }}
\end{align}
\begin{align}\notag
&{f_{ub}}\left( {{\rho ^j},{b^j}} \right) = f\left( {{\rho ^j},{b^j}} \right) + {{f'_\rho} }\left( {{\rho ^j},{b^j}} \right)\left( {\rho  - {\rho ^j}} \right)\\\tag{D.3}
& + {{f'_b}}\left( {{\rho ^j},{b^j}} \right)\left( {b - {b^j}} \right) + {{f''_{\rho \rho }}}\left( {{\rho ^j},{b^j}} \right){\left( {\rho  - {\rho ^j}} \right)^2},
\end{align}
\begin{align}\notag
&{g_{lb}}\left( {{\rho ^j},{a^j}} \right) = g\left( {{\rho ^j},{a^j}} \right) + {{g'_\rho} }\left( {{\rho ^j},{a^j}} \right)\left( {\rho  - {\rho ^j}} \right)\\\tag{D.4}
 &+ {{g'_a}}\left( {{\rho ^j},{a^j}} \right)\left( {a - {a^j}} \right) + {{g''_{aa}}}\left( {{\rho ^j},{a^j}} \right){\left( {a - {a^j}} \right)^2},
\end{align}
\begin{align}\notag
{w_{lb}}\left( {{\rho ^j},{a^j}} \right) = w\left( {{\rho ^j},{a^j}} \right) + {{w'_\rho} }\left( {{\rho ^j},{a^j}} \right)\left( {\rho  - {\rho ^j}} \right)\\\tag{D.5}
 + {{w'_a}}\left( {{\rho ^j},{a^j}} \right)\left( {a - {a^j}} \right) + {{w''_{aa}}}\left( {{\rho ^j},{a^j}} \right){\left( {a - {a^j}} \right)^2}.
\end{align}
The equal signs in the above inequalities only hold when $\rho  = {\rho ^j}$, $a = {a^j}$ and $b = {b^j}$. At the local points ${\rho ^j}$, ${{a^j}}$ and ${{b^j}}$, both the functions $y\left( \rho  \right)$, $f\left( {\rho ,b} \right)$, $g\left( {\rho ,a} \right)$, $w\left( {\rho ,a} \right)$ and its corresponding bounds ${y_{lb}}\left( {{\rho ^j}} \right)$, ${f_{ub}}\left( {{\rho ^j},{b^j}} \right)$, ${g_{lb}}\left( {{\rho ^j},{a^j}} \right)$, ${w_{lb}}\left( {{\rho ^j},{a^j}} \right)$ have the identical gradients.

\ifCLASSOPTIONcaptionsoff
  \newpage
\fi
\bibliographystyle{IEEEtran}
\bibliography{refa}

\end{document}